\documentclass[epj]{svjour}
\usepackage{graphics}
\begin{document}
\title{Atmospheric aerosols at the Pierre Auger Observatory and environmental implications}

\author{Karim Louedec\inst{1,2}\thanks{email: karim.louedec@lpsc.in2p3.fr} for the Pierre Auger Collaboration\inst{3}\thanks{http://www.auger.org/archive/authors\_2012\_06.html}, and R\'emi Losno\inst{4}\thanks{email: remi.losno@lisa.u-pec.fr} 
}                     

\institute{Laboratoire de Physique Subatomique et de Cosmologie (LPSC), UJF--INPG, CNRS/IN2P3, Grenoble, France \and Laboratoire de l'Acc\'el\'erateur Lin\'eaire (LAL), Univ Paris Sud, CNRS/IN2P3, Orsay, France \and Observatorio Pierre Auger, Av.\ San Mart\'{\i}n Norte 304, 5613 {Ma\-lar\-g\"{u}e}, Argentina \and Laboratoire Inter-universitaire des Syst\`emes Atmosph\'eriques (LISA), Universit\'e Paris-Est Cr\'eteil, Universit\'e Paris Diderot, CNRS/INSU, Cr\'eteil, France}

\authorrunning{K Louedec and R Losno}

\date{Published in Eur. Phys. J. Plus: 29 August 2012 }

\abstract{
The Pierre Auger Observatory detects the highest energy cosmic rays. Calorimetric measurements of extensive air showers induced by cosmic rays are performed with a fluorescence detector. Thus, one of the main challenges is the atmospheric monitoring, especially for aerosols in suspension in the atmosphere. Several methods are described which have been developed to measure the aerosol optical depth profile and aerosol phase function, using lasers and other light sources as recorded by the fluorescence detector. The origin of atmospheric aerosols traveling through the Auger site is also presented, highlighting the effect of surrounding areas to atmospheric properties. In the aim to extend the Pierre Auger Observatory to an atmospheric research platform, a discussion about a collaborative project is presented.
} 
\maketitle

\section{Introduction}
\label{sec:intro}
The Pierre Auger Observatory~\cite{PAO_1,PAO_2} is the largest operating cosmic ray observatory ever built. It was designed to measure the flux, arrival directions and mass composition of cosmic rays from $10^{18}~$eV (electron-volt) to the very highest energies. When cosmic rays enter the atmosphere, they induce extensive air showers composed of secondary particles. Charged particles excite atmospheric nitrogen molecules, and these molecules then emit fluorescence light in the $300-400~$nm range~\cite{AIRFLY,ArquerosFY}. At the Pierre Auger Observatory, the atmosphere is used as a giant calorimeter, representing a detector volume larger than $30\,000~$km$^3$. To minimise as much as possible the systematic errors of the fluorescence measurements, atmosphere properties have to be continuously monitored~\cite{AugerATMON}. During the development of an extensive air shower, the production rate of fluorescence photons depends on the temperature, pressure and humidity of the air~\cite{BiancaFY,VazquezFY,DelphinePHIL}. Then, from their production point to the telescope, these photons can be scattered by molecules (by Rayleigh scattering) and/or atmospheric aerosols (by Mie scattering). Thus, to track atmospheric parameters, an extensive atmospheric monitoring system has been developed that covers the whole array.

\begin{figure}[t]
\centering
\resizebox{0.65\textwidth}{!}{%
\includegraphics{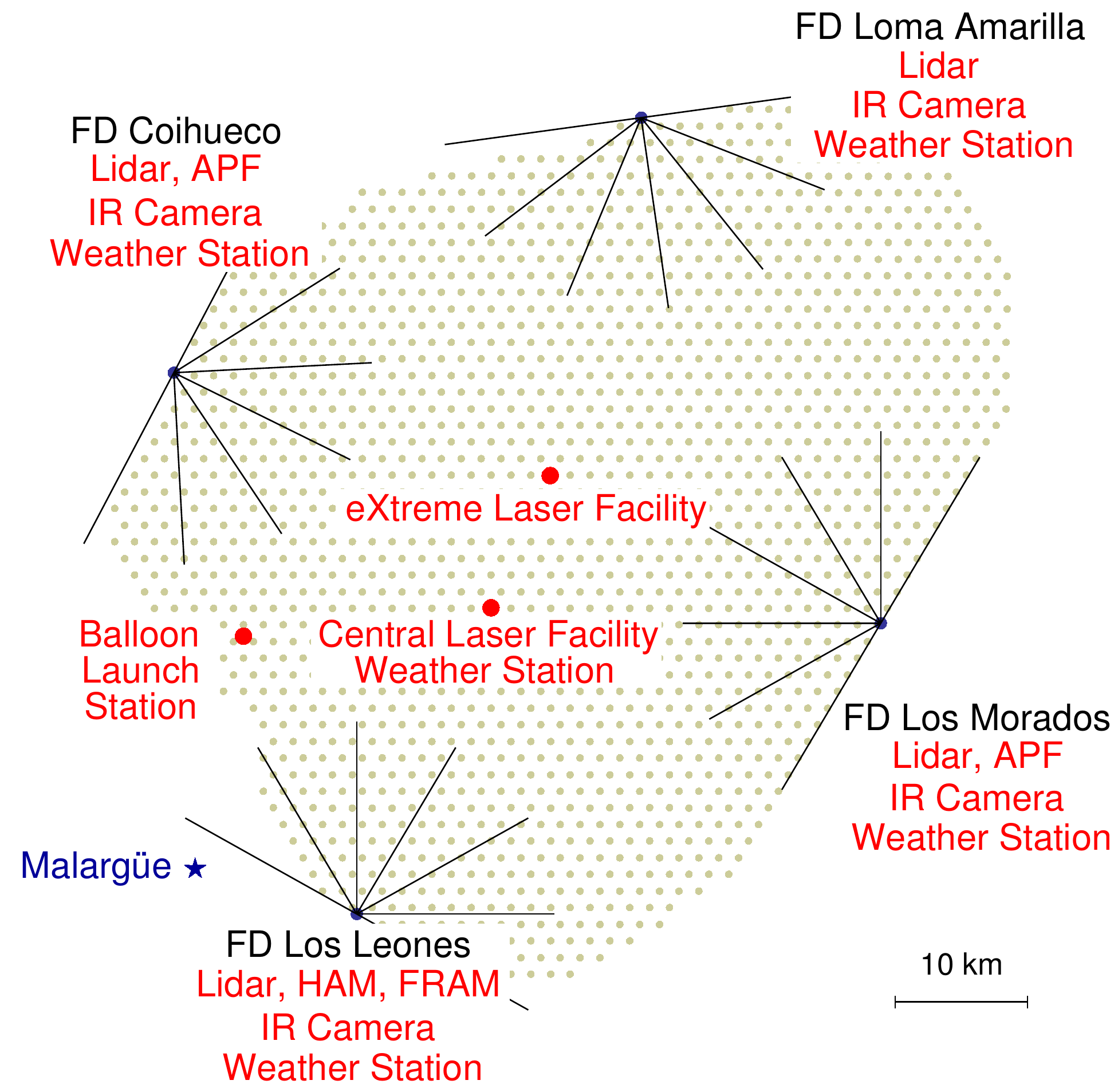}
}
\caption {{\bf Atmospheric monitoring map of the Pierre Auger Observatory (from~\cite{MyICRC}).} Gray dots show the positions of surface detector stations (SD). Black segments indicate the fields of view of the fluorescence detectors (FD) which are located in four sites, called Los Leones (LL), Los Morados (LM), Loma Amarilla (LA) and Coihueco (CO), on the perimeter of the surface array. Each FD site hosts several atmospheric monitoring facilities.}
\label{fig:MonitorArray}
\end{figure}

The different experimental facilities and their locations are shown in Fig.~\ref{fig:MonitorArray}. Atmospheric properties at ground level are provided by a network of five weather stations located at each fluorescence detector (FD) site and at the Central Laser Facility (CLF). They furnish atmospheric state variable measurements every five minutes. Also, meteorological radio-sonde flights with balloons have been operated (for more details, see~\cite{FocusPoint:BiancaMartin}). For the aerosol component, the central laser facility fires $50~$vertical shots every $15~$minutes during FD operations. Fluorescence telescopes, recording the Ultra-Violet (UV) laser tracks, are able to deduce the aerosol optical depth at different altitudes. In November 2008, a second laser facility called the eXtreme Laser Facility, or XLF, was deployed. These aerosol profile measurements are complemented by cloud measurements by four elastic backscattering lidars located at each eye (for more details, see~\cite{FocusPoint:lidar}). A Raman lidar currently under test in Colorado (USA) is scheduled to be moved to the Auger Observatory for the Super-Test-Beam project~\cite{WienckeICRC}. To improve our knowledge of photon scattering on aerosols, two Aerosol Phase Function monitors (APF) have been installed at the Coihueco and Los Morados FD sites. The APF instruments generate a collimated horizontal light beam produced by a Xenon flasher. The light passes in front of one FD site. The aerosol attenuation depends on the incident wavelength. This measurement is the main goal of two optical telescopes in Auger, the Horizontal Attenuation Monitor, or HAM~\cite{HAM_proceeding}, and the (F/ph)otometric Robotic Telescope for Atmospheric Monitoring, or FRAM~\cite{FRAM_proceeding}. 

\section{Aerosol effects on light propagation in the atmosphere}
\label{sec:aerosol_effects}
Although the atmosphere is mainly composed of molecules, small particles such as dust or droplets may be in suspension. These particles are called {\it atmospheric aerosols} and their typical size varies from a few nanometres to a few micrometres. Most of the atmospheric aerosols are present only in the first few kilometres above the ground. Unlike the molecular component, the aerosol population is highly variable in time and location, depending on the wind and weather conditions. Aerosol conditions affect the propagation of UV light from air showers. Even if the absolute fluorescence yield remains the largest source of uncertainty for FD measurements~\cite{PRL_spectrum}, aerosol effects also contribute significantly to systematic uncertainties. Absorption in the air is usually negligible for photons having a wavelength in the UV domain. Thus, attenuation will represent only the scattering phenomena. The main atmospheric attenuation processes are Rayleigh scattering, by the molecular component, and Mie scattering, by the aerosol component, both being elastic. In the following, only scattering on aerosols will be discussed. For more details on the molecular component of the atmosphere, see~\cite{FocusPoint:BiancaMartin}.

Two main physical quantities have to be estimated to correct the effect of the aerosols on the number of photons detected by the telescopes. These are the aerosol attenuation length as a function of height, linked to the aerosol optical depth, and the aerosol scattering phase function.

\subsection{Aerosol transmission -- Mie regime}
\label{sec:transmission_mie}
Aerosol attenuation of light from an air shower to the FD telescope can be expressed as a transmission coefficient $\Gamma_{\rm a}$, or optical transmittance, giving the fraction of incident light at a specified wavelength $\lambda$ that passes through an atmosphere of thickness $x$. If $\tau_{\rm a}$ is the aerosol optical depth, then $\Gamma_{\rm a}$ is estimated using the Beer-Lambert law
\begin{equation}
\Gamma_{\rm a}(x,\lambda) = \exp\,\left[-\tau_{\rm a}(x,\lambda)\right],
\end{equation}
under the assumption of horizontal uniformity. Usually, the aerosol population can be described as a superposition of several horizontally uniform layers, so the spatial dependence of the aerosol optical depth can be reduced to an altitude dependence. Hence, the aerosol transmission from an altitude $h$ to the ground through a slanted path of zenith angle $\theta$ is
\begin{equation}
\Gamma_{\rm a}(h,\theta,\lambda) = (1 + H.O.) \, \exp\,\left[-\frac{\tau_{\rm a}(h,\lambda)}{\cos\theta}\right],
\end{equation}
where $H.O.$ represents a higher-order correction that accounts for the single and multiple scattering of photons into the field of view of the telescope.

Therefore, knowledge of the aerosol transmission parameters at the Pierre Auger Observatory will be acquired through measurements of the aerosol optical depth $\tau_{\rm a}(h,\lambda)$ at different location in the Auger array, throughout the night. Usually, the aerosol concentration decreases rapidly with the altitude. The aerosol optical depths are measured in the field at a fixed wavelength $\lambda_0$. To evaluate the aerosol extinction for a given incident wavelength, a common parameterization used is a power law due to \r{A}ngstr\"om,
\begin{equation}
\tau_{\rm a}(h,\lambda) = \tau_{\rm a}(h,\lambda_0)\, \left( \frac{\lambda_0}{\lambda}\right)^\gamma,
\label{eq:transmission_mie_1}
\end{equation}
where $\gamma$ is known as the \r{A}ngstr\"om coefficient. This exponent depends on the size distribution of the aerosols. When the aerosol particle size approaches the size of air molecules, $\gamma$ should tend to $4$ (mainly dominated by {\it accumulation-mode} aerosols), and for very large particles, typically larger than $1~\mu$m, it should approach zero (dominated by {\it coarse-mode} aerosols). Usually, a $\gamma \simeq 0$ is characteristic of a desert environment and the aerosol optical depth is more or less independent of the wavelength. Measurements at the Pierre Auger Observatory gave $\gamma$ values close to zero (see Sect.~\ref{sec:wLength_dependence}).

\subsection{Angular dependence of aerosol scattering}
\label{sec:angular_dependence}
Due to the collecting area of the FD telescopes and the typical distance between a telescope and an air shower, only a small fraction of the photons produced in the air shower is detected without scattering. Thus, the scattering properties of the atmosphere need to be well estimated. The angular dependence of the scattering may be described by a phase function $P(\zeta)$, defined as the probability per unit solid angle for scattering out of the beam path through an angle $\zeta$. Following the convention for the atmospheric domain, ${\sigma_{\rm a}}^{-1}\,{\rm d} \sigma_{\rm a}/{\rm d} \Omega$ is the normalised differential aerosol scattering cross section, which is identical to the aerosol phase function $P_{\rm a}(\zeta)$. The integral of $P_{\rm a}(\zeta)$ over all solid angles has to be equal to unity.

\begin{figure}[!t]
\centering
\resizebox{1.0\textwidth}{!}{%
\includegraphics{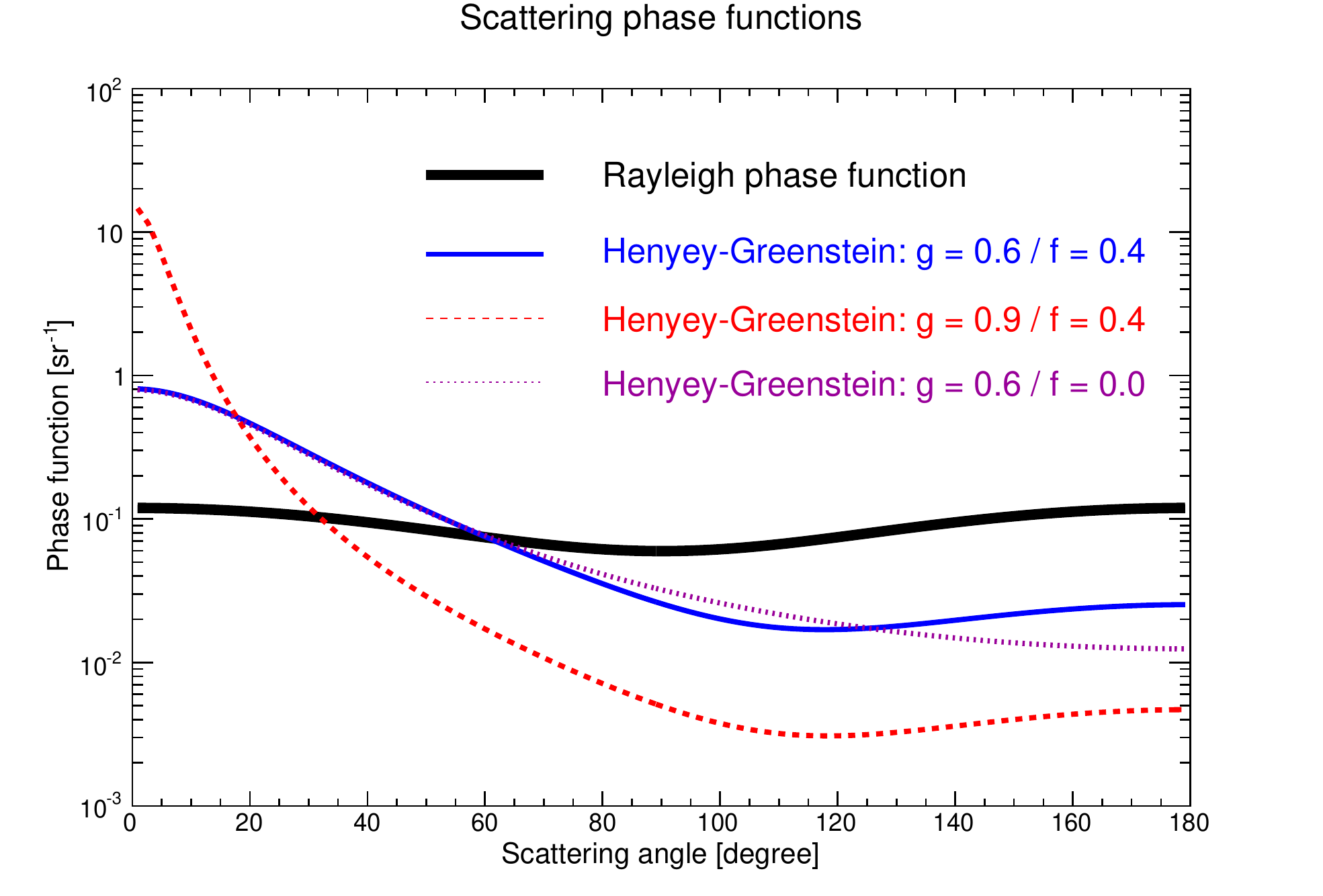}
}
\caption {{\bf Henyey-Greenstein functions representing the aerosol phase function for different asymmetry parameters g$_{\rm \bf HG}$ and backward factors f.} The Rayleigh phase function, proportional to $(1+\cos^2 \zeta)$ and representing scattering properties for the molecular component of the atmosphere, is also plotted.}
\label{fig:PhaseFunctionHG}
\end{figure}

Whereas the molecular component is described analytically by the Rayleigh scattering theory~\cite{Bucholtz1995}, the Mie scattering cannot be described by a basic equation for the aerosol component. This is due to the fact that the scattering cross section depends on the size distribution and shape of the scattering centres. Typically, forward scattering dominates in the Mie regime and the forward-backward ratio can vary strongly with aerosol type. More details on the relationship between scattering phase function and aerosol size can be found in~\cite{MyRamsauer} where a phenomenological approach is developed. The aerosol phase function used by the Pierre Auger Collaboration is parameterised by a modified Henyey-Greenstein function~\cite{HenyeyG}
\begin{equation}
P_{\rm a}(\zeta|g,f) = \frac{1-g^2}{4\pi}\left[\frac{1}{\left(1 + g^2 - 2\,g\,\cos\zeta \right)^{3/2}} + f\,\frac{3\cos^2\zeta-1}{2\,\left(1+g^2 \right)^{3/2}}  \right],
\label{eq:angular_dependence_2}
\end{equation}
where $g= g_{\rm HG} = \langle \cos\zeta \rangle$ is the asymmetry parameter and $f$ the strength of the second component to the backward scattering peak ($f=0$ meaning no additional component). The asymmetry parameter provides the scattered light intensity in the forward direction: a larger $g$ means more forward-scattered light (see Fig.~\ref{fig:PhaseFunctionHG}). Values go from $g~=~1$ (pure forward scattering) to $g~=~-1$ (pure backward scattering), with $g~=~0$ meaning isotropic scattering. The second term in the expression -- a second-order Legendre polynomial, chosen not to affect the normalization of the phase function -- is introduced to describe the extra backscattering component. At the Pierre Auger Observatory, the goal in monitoring the aerosol phase function is to estimate the $\lbrace g,f\rbrace$ parameters, two observable quantities depending on local aerosol properties (see Sect.~\ref{sec:aerosol_scat_measurements}).


\section{Aerosol measurements at the Pierre Auger Observatory}
\label{sec:aerosol_ measurements}
Several instruments are deployed at the Pierre Auger Observatory to observe aerosol scattering properties. The aerosol optical depth is estimated using UV laser measurements from the central lasers and scanning lidars (more details on lidar measurements in~\cite{FocusPoint:lidar}); the aerosol phase function is determined with Aerosol Phase Function monitors and the wavelength dependence of the aerosol optical depth is measured with the optical telescopes HAM and FRAM. In addition, an aerosol sampling program based on filters was undertaken from June to November 2008.

\subsection{Aerosol optical depth measurements}
\label{sec:vaod_ measurements}
At the Pierre Auger Observatory, the aerosol optical depth is measured through the night using the fluorescence detector to measure light from the central laser facility, located on-site towards the centre of the SD. The main role of the CLF is to produce calibrated laser ``test beams''. It is powered by a battery, charged by solar panels~\cite{CLF_jinst}. The main component is a laser with a wavelength fixed at $355~$nm, in the middle of the nitrogen fluorescence spectrum produced by air showers~\cite{AIRFLY}. The pulse width of the beam is $7$~ns and a maximum energy per pulse is around $7$~mJ. This is of the order of fluorescence light produced by a shower with an energy of $10^{20}~$eV. To estimate the relative energy of each laser pulse, a portion of the beam is diverted into a photo-diode detector. The analysis of the vertical aerosol optical depth uses a vertical beam. A steering head mounted on the roof of the container makes it possible to direct the beam towards any direction above the horizon.



\begin{figure}[!t]
\centering
\resizebox{1.0\textwidth}{!}{%
\includegraphics{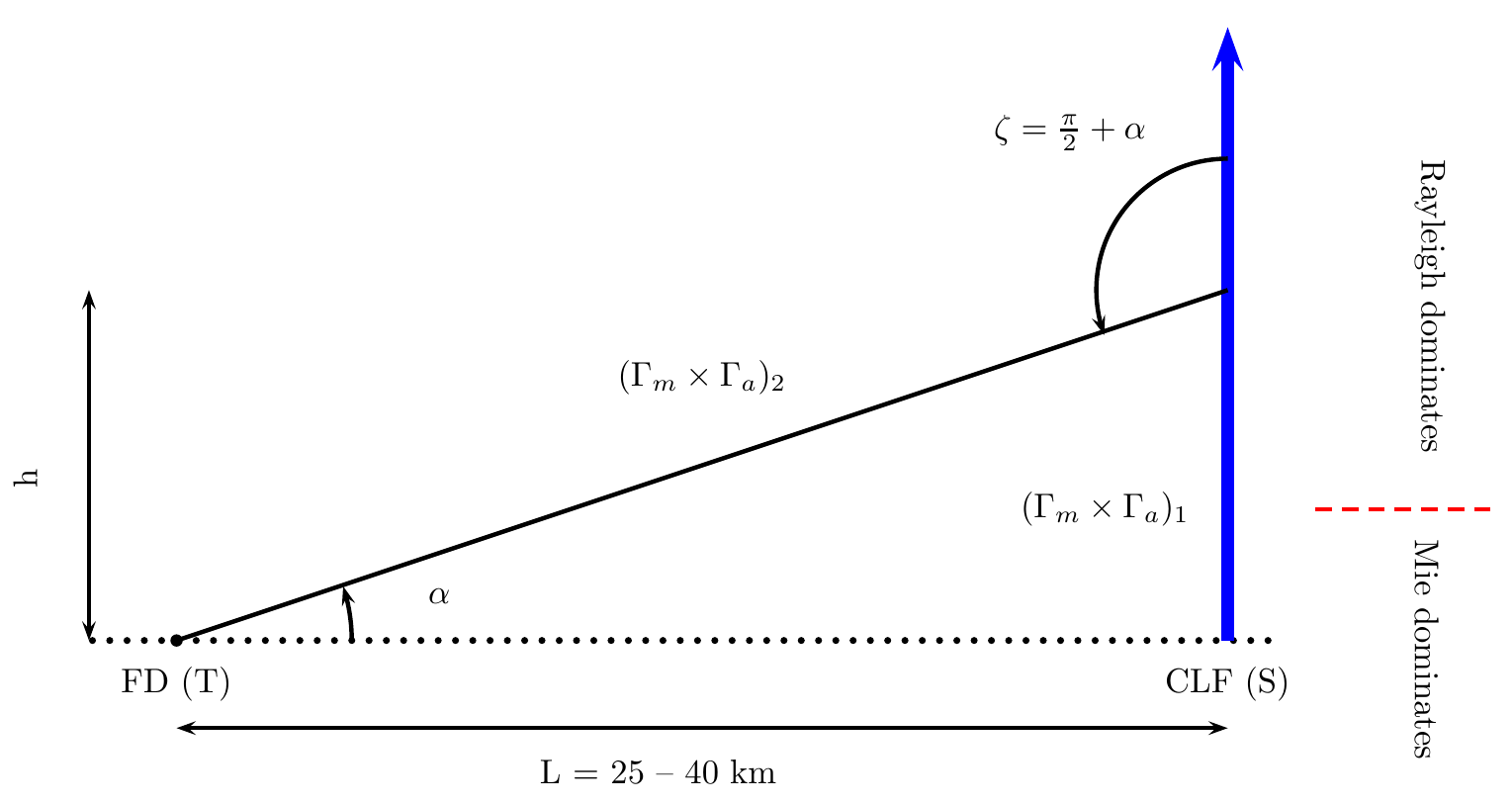}
}
\caption{{\bf Geometrical arrangement, viewed from the side, of the central laser CLF and the FD telescope.} The light is scattered out of the laser beam at a height $h$ corresponding to an elevation angle $\alpha$ and a scattering angle $\zeta = \pi/2+\alpha$. $(\Gamma_{\rm m}\;\Gamma_{\rm a})_1$ and $(\Gamma_{\rm m}\;\Gamma_{\rm a})_2$ are the total attenuations from the CLF to the scattering location and from the scattering location to the FD, respectively. The transition between the Mie domination and the Rayleigh domination is located just a few kilometres Above Ground Level (AGL), dependent on the aerosol conditions.}
\label{fig:CLF_geometry}
\end{figure}

\begin{figure}[!t]
\centering
\resizebox{1.0\textwidth}{!}{%
\includegraphics{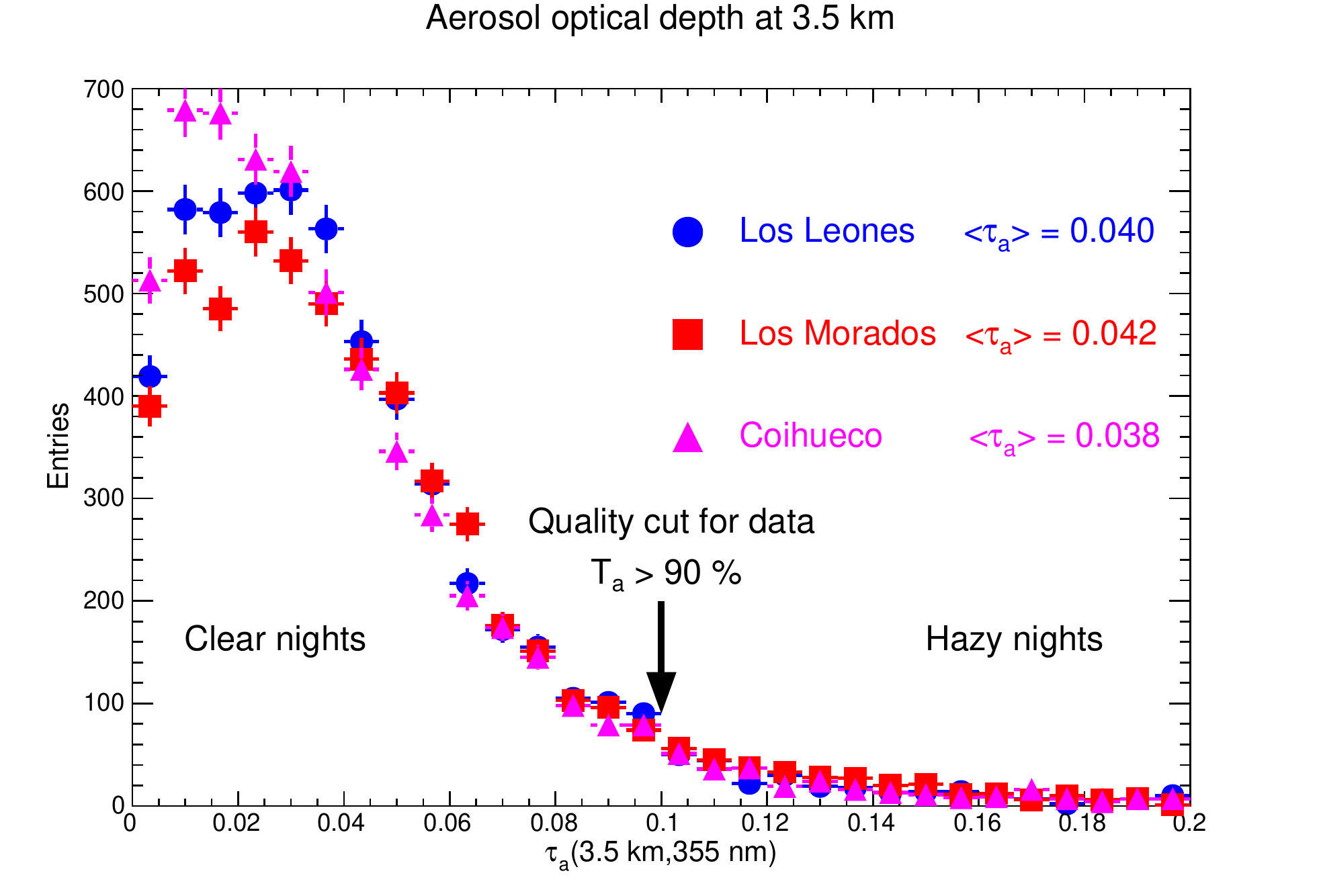}
}
\caption{{\bf Aerosol optical depth at ${\bf 3.5~}$km above the fluorescence telescopes at Los Leones, Los Morados and Coihueco.} The wavelength of the CLF laser is $355~$nm. Data were acquired between January 2004 and December 2010.}
\label{fig:CLF_VAOD_values}
\end{figure}

When a laser shot is fired, the FD telescopes collect a small fraction of the light scattered out of the laser beam. The recorded signal is not constant but depends on the atmosphere properties. Thus, a method has been developed in the Auger Collaboration to estimate the vertical aerosol optical depth $\tau_{\rm a}(h,\lambda_0)$ with the CLF, where $\lambda_0$ is the CLF wavelength~\cite{LauraValore}. The method to determine $\tau_{\rm a}(h,\lambda_0)$ normalises the measurement of the laser beam to the signal that would be recorded under aerosol-free atmospheric conditions. The molecular references are average CLF laser profiles measured during very clear nights, the so-called ``reference clear nights''. Once a light profile, averaged over one hour, is normalised using a clear-night reference, the attenuation of the remaining light is due to aerosol scattering along the path from the laser beam to the FD telescope. A horizontal uniformity for the molecular and aerosol components is assumed. The amount of light from the laser beam reaching the detector at the elevation angle $\alpha$ is written as
\begin{equation}
N_{\rm obs}(\alpha) = N_{\rm 0} \, \left(\Gamma_{\rm m}\;\Gamma_{\rm a}\right)_1\, \left[P_{\rm m}\left(\frac{\pi}{2}+\alpha \right) + P_{\rm a} \left(\frac{\pi}{2}+\alpha \right)\right]\, \left(\Gamma_{\rm m}\;\Gamma_{\rm a}\right)_2,
\label{eq:clf1}
\end{equation}
where $N_{\rm 0}$ is the number of photons produced per laser pulse, $\lbrace\Gamma_{\rm m},\Gamma_{\rm a}\rbrace$ are the molecular and aerosol transmission factors, and $\lbrace P_{\rm m},P_{\rm a}\rbrace$ are the molecular and aerosol phase functions (see Fig.~\ref{fig:CLF_geometry}). The indices $1$ and $2$ correspond to the way from the CLF to the scattering location, and from the scattering location to the detector.

In the case of a clear night, the aerosol transmission factor $\Gamma_{\rm a}$ is equal to one and the scattering over aerosols is negligible. Eq.~(\ref{eq:clf1}) is reduced to molecular part only
\begin{equation}
N_{\rm mol}(\alpha) = N_{\rm 0} \, \left(\Gamma_{\rm m}\right)_1\, \left[P_{\rm m}\left(\frac{\pi}{2}+\alpha \right)\right]\, \left(\Gamma_{\rm m}\right)_2.
\end{equation}
~\\

Applying the aerosol horizontal uniformity condition, $\Gamma_{\rm a}(h,\alpha,\lambda_0) = \exp\,\left[-\tau_{\rm a}(h,\lambda_0)/\sin\alpha \right]$, and combining $N_{\rm obs}$ and $N_{\rm mol}$, Eq.~(\ref{eq:clf1}) becomes
\begin{equation}
\tau_{\rm a}(h,\lambda_0) = -\frac{\sin\alpha}{1+\sin\alpha}\left[{\rm ln}\left(\frac{N_{\rm obs}(\alpha)}{N_{\rm mol}(\alpha)}\right) - {\rm ln}\left(1 + \frac{P_{\rm a} (\frac{\pi}{2}+\alpha)}{P_{\rm m} (\frac{\pi}{2}+\alpha)}\right) \right].
\label{eq:clf2}
\end{equation}

In the scattering angle range seen by the FD telescopes (between $90^{\circ}$ and $120^{\circ}$ here) the aerosol scattering contribution is much lower than the molecular contribution. Eq.~(\ref{eq:clf2}) simplifies to
\begin{equation}
\tau_{\rm a}(h,\lambda_0) = \frac{\sin\alpha}{1+\sin\alpha}\,{\rm ln}\left(\frac{N_{\rm mol}(\alpha)}{N_{\rm obs}(\alpha)}\right).
\label{eq:clf3}
\end{equation}

With these approximations, the aerosol optical depth formula depends only on the elevation angle $\alpha$ of each laser track segment, linked to the altitude by $h =L\, \tan\alpha + h_{\rm Auger}$, where $L$ is the horizontal distance between the FD site and the CLF, and  $h_{\rm Auger}$ the altitude of the Auger array above sea level.

\begin{figure}[!t]
\centering
\resizebox{0.49\textwidth}{!}{%
\includegraphics{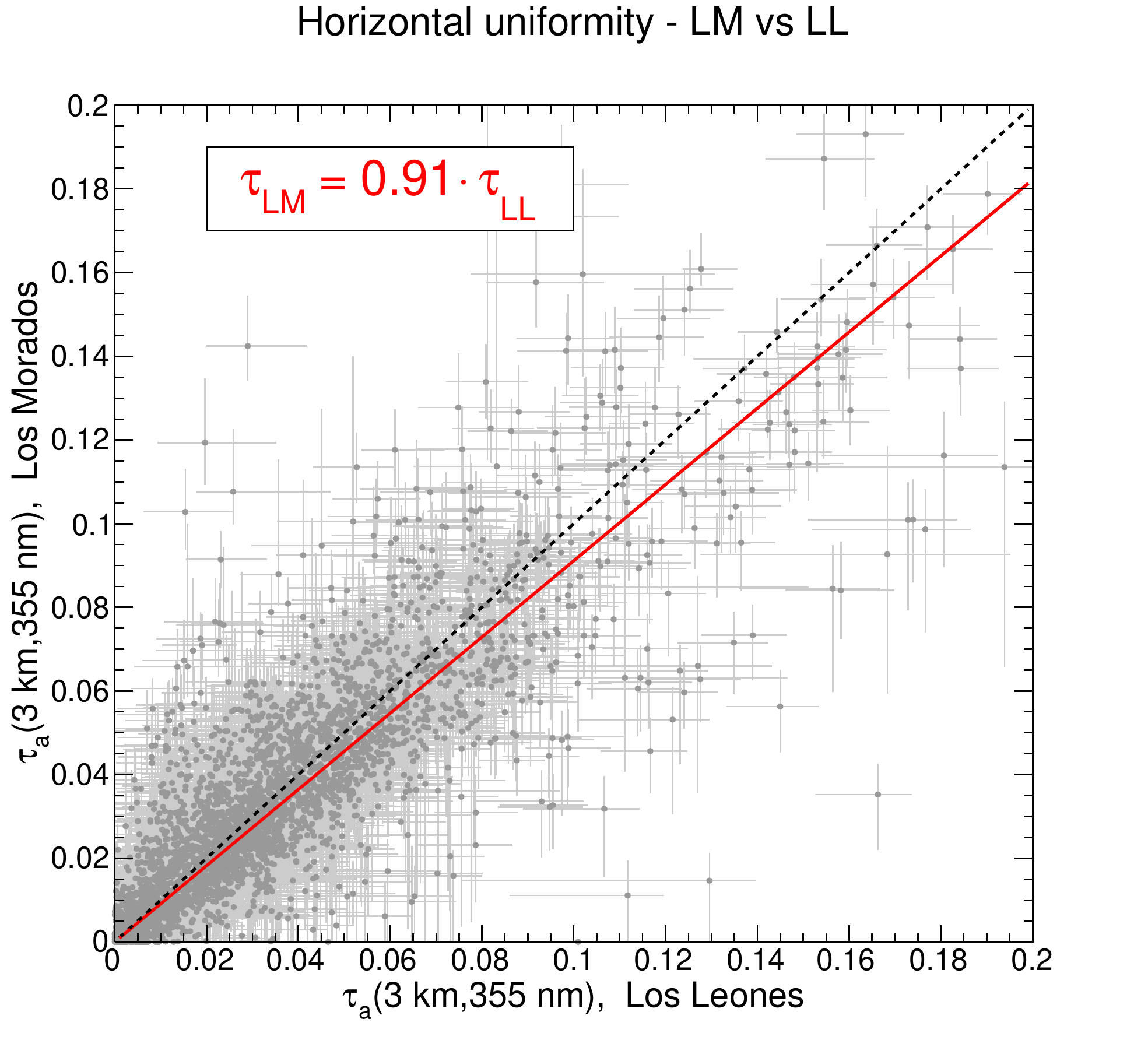}
}
\resizebox{0.49\textwidth}{!}{%
\includegraphics{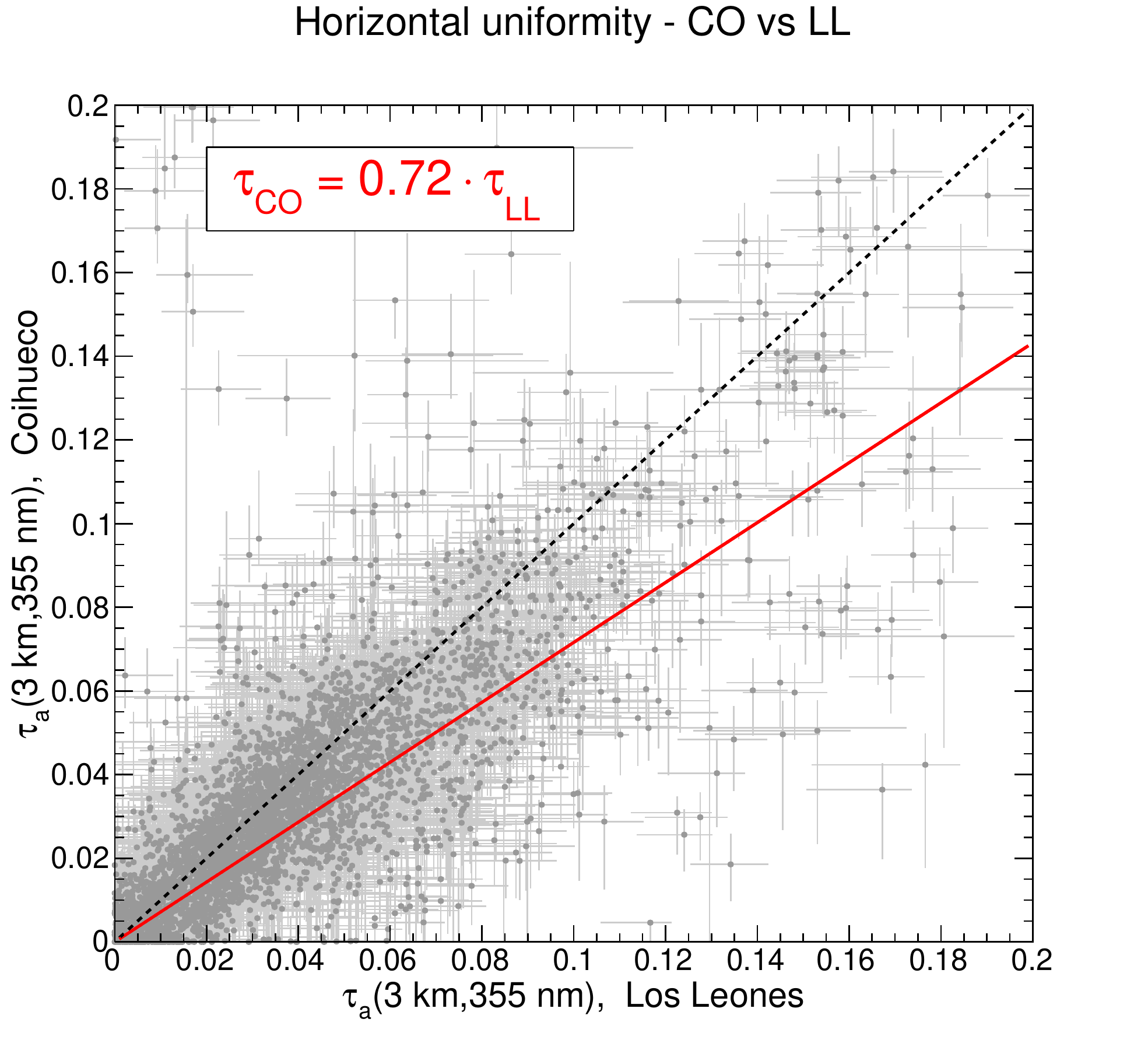}
}
\caption{{\bf Horizontal uniformity of the aerosol optical depths measured with CLF shots at Los Leones (LL), Los Morados (LM) and Coihueco (CO).} Scatter plots show the correlation between two FD sites, the dashed line representing full agreement between two data points and the solid line the fit to data (from~\cite{AugerATMON}).}
\label{fig:CLF_VAOD_HU}
\end{figure}

Figure~\ref{fig:CLF_VAOD_values} depicts the aerosol optical depth distribution recorded at Los Leones, Los Morados and Coihueco between 2004 and 2010. Measurements from Loma Amarilla are not available due to its large distance from the CLF site. The XLF, closer to Loma Amarilla, provides optical depth information for this site. Typically, at $3.5~$km above the FD level, the mean value for $\tau_{\rm a}(h,\lambda_0)$ is around 0.04. Note that a cut at $0.1$ is used in the air shower reconstruction as a quality cut: all the events occurring in a night with a $\tau_{\rm a}(h,\lambda_0) \geq 0.1$ are rejected.

Horizontal uniformity is usually assumed for the molecular component. Figure~\ref{fig:CLF_VAOD_HU} shows the scatter plots of the aerosol optical depth $\tau_{\rm a}(h,\lambda_0)$ measured at Los Morados and Coihueco with respect to the ones measured at Los Leones. The altitude is fixed at $3~$km above the FD level and the wavelength at $355~$nm (CLF wavelength). This means that different altitudes are probed at Los Leones and Coihueco, the latter being almost $300~$m higher. The optical depth data for the different sites suggest that aerosol conditions differ with location. The agreement is better between Los Leones and Los Morados, two sites differing only slightly in altitude. The difference could be also explained by a different composition of the soil between the Coihueco site and the other ones, aerosols coming mainly from the ground. To take into account this non-uniformity, the implementation of the aerosol parameters into the Auger Offline software divides the Auger array into five zones centred on the midpoints between the FD buildings and the CLF. Horizontal uniformity of the aerosol component is assumed in each slice. Then, each region is divided vertically into layers, each layer having a thickness equal to $200~$metres.

\subsection{Aerosol scattering measurements}
\label{sec:aerosol_scat_measurements}
The FD reconstruction of the cosmic ray energy has to take into account not only the light attenuated during propagation, but also has to remove the multiple scattering component adding to the fluorescence light contamination. Aerosol scattering is described by the aerosol phase function $P_{\rm a}(\zeta)$ which can be parameterised by a modified Henyey-Greenstein function as in Eq.~(\ref{eq:angular_dependence_2}). At the Pierre Auger Observatory, two Aerosol Phase Function monitors, in conjunction with the FD telescopes, are used to measure the parameters $\lbrace g,f\rbrace$ on an hourly basis during FD data acquisition. The APF light sources emit a near-horizontal pulsed light beam in the field of view of their nearby FD site at Coihueco and Los Morados. Each APF building contains collimated Xenon flash lamp sources, firing an hourly sequence of $350~$nm and $390~$nm shots. The aerosol phase function is then reconstructed from the intensity of the light observed by the FD cameras as a function of scattering angle $\zeta$, for angles between $30^{\circ}$ and $150^{\circ}$. After corrections for geometry, attenuation and collection efficiency for each pixel, the binned APF signal $S(\zeta)$ observed is subjected to a 4-parameter fit

\begin{equation}
S(\zeta) = C\, \left[\frac{1}{\Lambda_{\rm m}(h_{\rm Auger})}\, P_{\rm m}(\zeta) + \frac{1}{\Lambda_{\rm a}(h_{\rm Auger})}\, P_{\rm a}(\zeta |g,f)   \right],
\label{eq:clf3}
\end{equation}
where $\lbrace C/\Lambda_{\rm m}(h_{\rm Auger}),C/\Lambda_{\rm a}(h_{\rm Auger}),g,f\rbrace$ are the fit parameters and $P_{\rm m}(\zeta)$ the Rayleigh phase function~\cite{BenZviEtAl}. The first two fit parameters can be used to estimate the molecular attenuation length and the aerosol attenuation length, respectively; while $g$ and $f$ are used to estimate the aerosol size distribution.

\begin{figure}[!t]
\centering
\resizebox{0.49\textwidth}{!}{%
\includegraphics{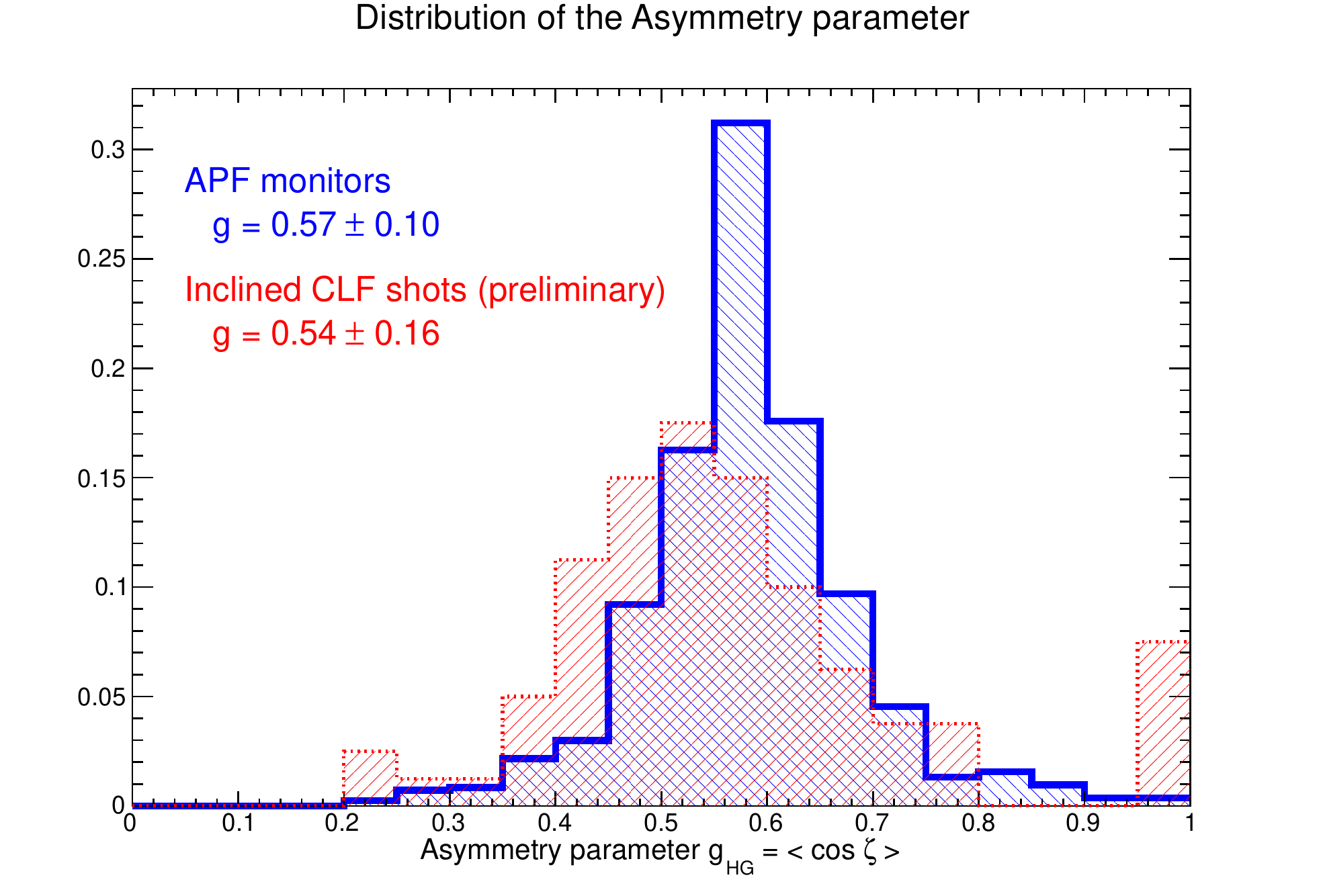}
}
\resizebox{0.49\textwidth}{!}{%
\includegraphics{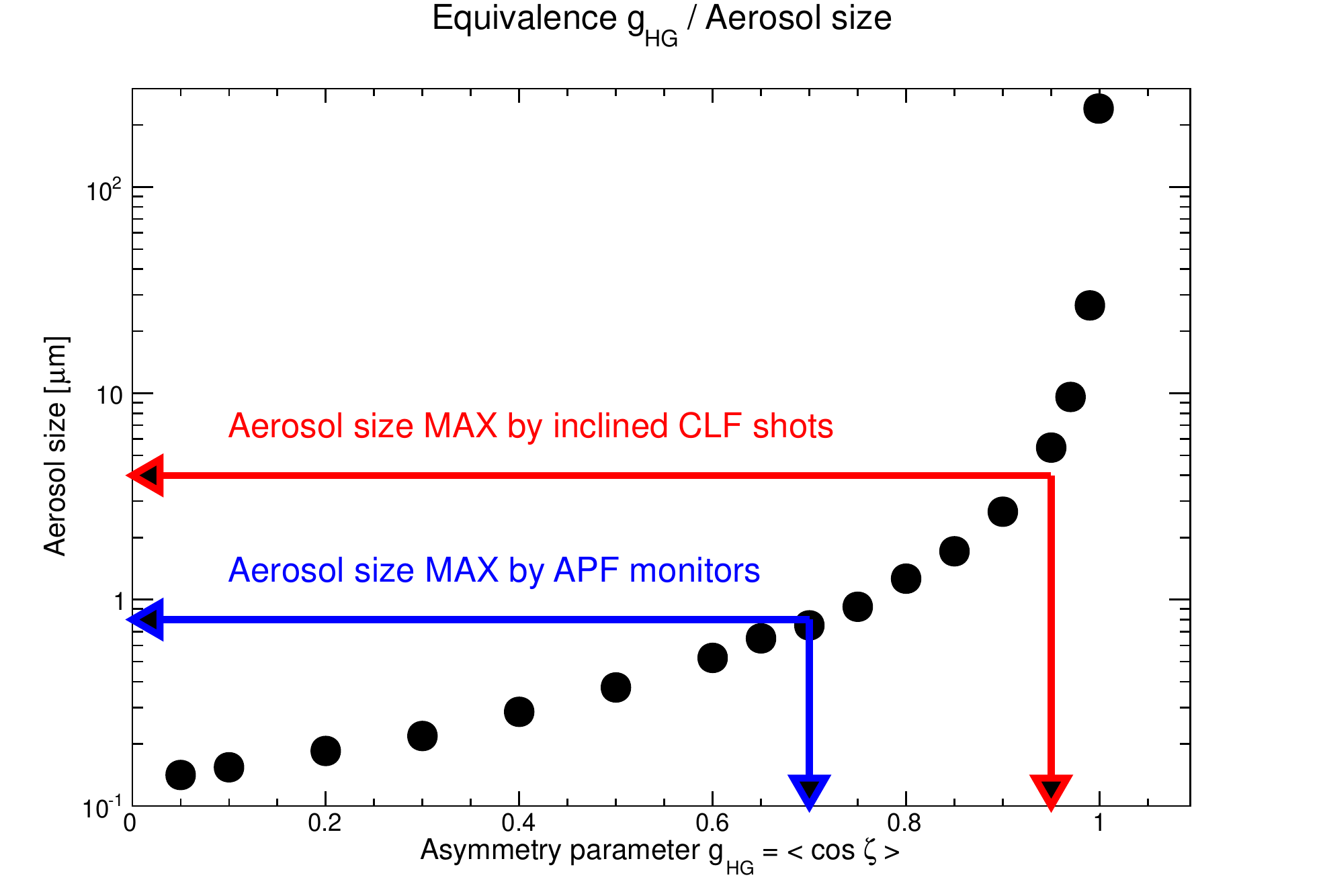}
}
\caption{{\bf Aerosol size measurements by the APF monitors and the very inclined CLF shots.} {\it Left panel}: Asymmetry parameter distribution measured by the APF monitor {\it (continuous line)} and the very inclined CLF laser shots {\it (dotted line)}. The APF monitor data were measured at Coihueco between June 2006 and July 2008~\cite{AugerATMON}. The values from the new method were obtained with laser shots at Los Leones recorded in 2008~\cite{MyThesis}. {\it Right panel}: Relationship between the asymmetry parameter of the modified Henyey-Greenstein function and the mean aerosol size, {\it i.e.}\ the diameter in the case of spherical aerosols. Upper limits on the aerosol size are plotted for the two methods, following the Ramsauer approach~\cite{MyRamsauer}.}
\label{fig:gFactorDISTRIBUTION}
\end{figure}

Figure~\ref{fig:gFactorDISTRIBUTION}~(left) shows the distribution for the asymmetry parameter measured at Coihueco between June 2006 and July 2008. The average $g$ value is $\langle g\rangle = 0.57 \pm 0.10$, when one excludes the $g=0$ nights corresponding to aerosol-free nights. Such a value for the asymmetry parameter corresponds to values usually measured in desert locations with significant levels of sand and soil dust~\cite{LongtinMilitaryReport}. The second parameter of the modified Henyey-Greenstein function, the backscatter coefficient, is estimated as $\langle f\rangle = 0.40 \pm 0.10$. It means that a backward peak for the aerosol phase function exists. Hence, at the Pierre Auger Observatory, the typical aerosol phase function is described by $P_{\rm a}(\zeta |g = 0.6,\,f = 0.4)$.

Recently, a new method based on very inclined shots fired by the CLF was developed (laser shots with zenith angles higher than $86^\circ$)~\cite{MyThesis}. Following the same idea as previously, knowing the geometry of the laser shot and the signal recorded by the pixels, it is possible to extract the $g$ parameter. The advantage of this technique is that a $g$ parameter can be estimated for each FD site, and it can cover lower scattering angles as shown in Fig.~\ref{fig:gFactorDISTRIBUTION}~(right) (the angular range where larger aerosols could be detected). It gives a similar estimation for the asymmetry parameter: $\langle g\rangle = 0.54 \pm 0.16$ (see Fig.~\ref{fig:gFactorDISTRIBUTION}~(left)).

\subsection{Wavelength dependence}
\label{sec:wLength_dependence}

\begin{figure}[!b]
\centering
\resizebox{1.\textwidth}{!}{%
\includegraphics{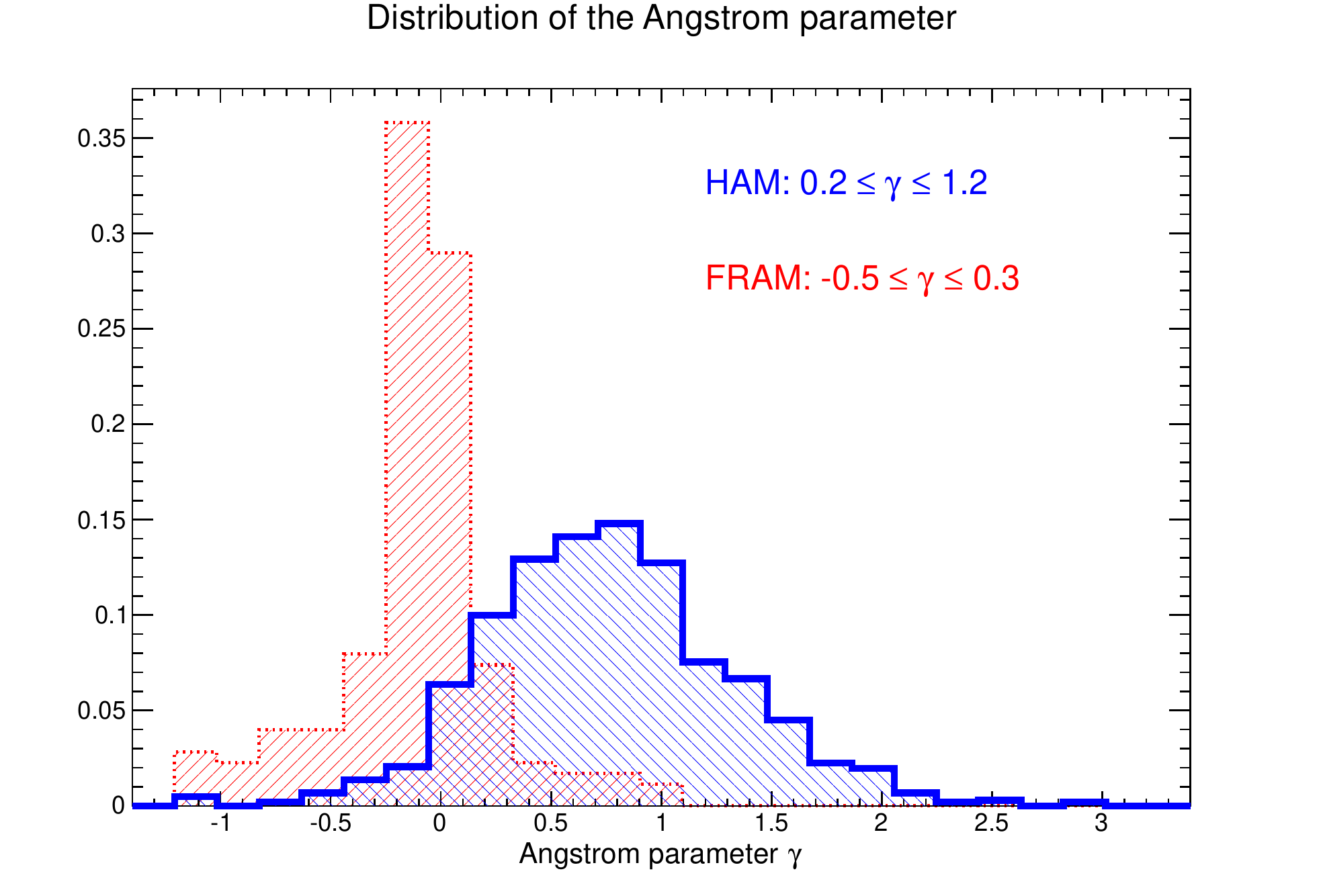}
}
\caption{{\bf Normalised distributions of the \r{A}ngstr\"om exponent observed by the two monitors dedicated to this measurement at the Pierre Auger Observatory.} Estimations of $\gamma$ from the HAM for data recorded between July 2006 and February 2007 {\it (continuous line)}~\cite{HAM_proceeding}. The measured $\gamma$ distribution for data collected by the FRAM from June 2006 until December 2008 {\it (dotted line)}~\cite{FRAM_proceeding}.}
\label{fig:Angstrom_parameter}
\end{figure}

Up to now, the measurements to estimate the aerosol optical depth have been done at a fixed wavelength, that of the laser of the CLF. The empirical parameterization defined in Eq.~(\ref{eq:transmission_mie_1}) gives a basic way to get the aerosol effect for any wavelength if the \r{A}ngstr\"om exponent is known. At the Pierre Auger Observatory, two different facilities were installed to estimate this key parameter: the Horizontal Attenuation Monitor (HAM) and the F/(Ph)otometric Robotic Atmospheric Monitor (FRAM).

The HAM consists of a high intensity discharge lamp located at Coihueco, providing a large wavelength range source. To record the emitted light, a CCD camera is placed around $45~$km away at Los Leones~\cite{HAM_proceeding}. With this configuration, the total horizontal atmospheric attenuation across the Auger array can be obtained. Using a filter wheel, the camera records the aerosol attenuation between the two sites at five different wavelengths between $350$ and $550~$nm. By fitting the signal with respect to the wavelength, it is possible to estimate the \r{A}ngstr\"om exponent $\gamma$. From the data recorded between July 2006 and February 2007, the wavelength dependence displays a value $0.2 \leq \gamma \leq 1.2$ (see Fig.~\ref{fig:Angstrom_parameter}). The uncertainties are dominated by measurement fluctuations, and include a systematic effect due to subtraction of the estimated molecular attenuation between Los Leones and Coihueco.

The main task of the FRAM telescope is the continuous monitoring of the wavelength dependence of the total column aerosol optical depth~\cite{FRAM_proceeding}. The facility is located close to the Los Leones FD building. The method is the following: the telescope observes a set of chosen standard stars and, from these observations, it obtains extinction coefficients and their wavelength dependence. Five filters are used, for wavelengths between $360$ and $547~$nm. As preliminary results, the FRAM telescope estimates a value $-0.5 \leq \gamma \leq 0.3$ (see Fig.~\ref{fig:Angstrom_parameter}), lower than the results from the HAM instrument.

The small value of the exponent $\gamma$ suggests that the Auger Observatory has a large component of coarse-mode aerosols, meaning aerosols larger than around $1~\mu$m.

\subsection{Aerosol sampling measurements}
\label{sec:aerosol_sampling_measurements}

\begin{figure}[!t]
\centering
\resizebox{0.39\textwidth}{!}{%
\includegraphics{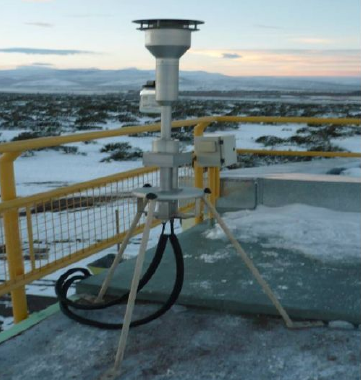}
}
\resizebox{0.6\textwidth}{!}{%
\includegraphics{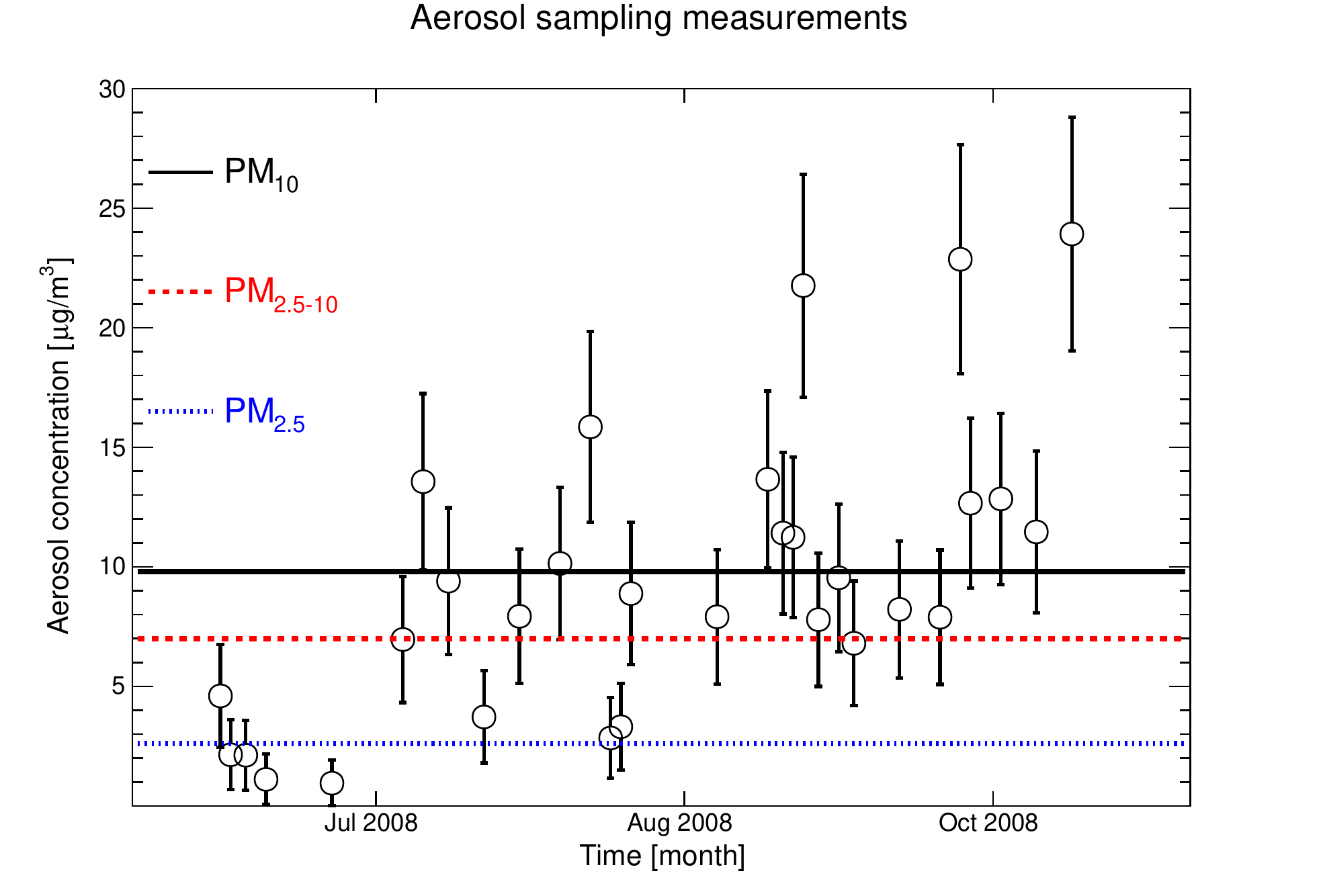}
}
\caption{{\bf Aerosol sampling at the Coihueco site (from~\cite{MariaI_LongPaper}).} {\it Left panel}: Andersen-Graseby 240 dichotomous sampler located at Coihueco, on the roof of the FD building. {\it Right panel}: Aerosol mass concentration for PM$_{\rm 10}$ measurements at the Coihueco sampling site (the circles). The horizontal lines give the mean values for the three aerosol samplings.\label{fig:AndersenGraseby}}
\end{figure}

\begin{figure}[!t]
\centering
\resizebox{0.49\textwidth}{!}{%
\includegraphics{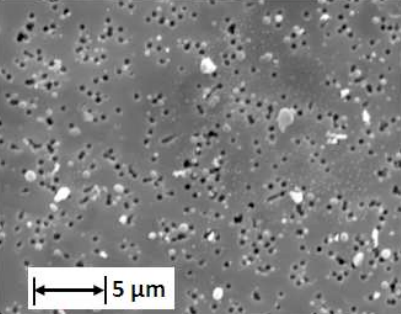}
}
\resizebox{0.49\textwidth}{!}{%
\includegraphics{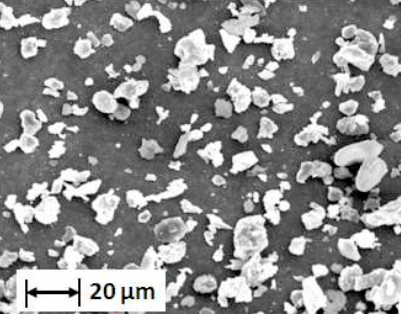}
}
\caption{{\bf Micrograph using Scanning Electron Microscopy (from~\cite{WienckeICRC}).} {\it Left panel: }Sampling of PM$_{\rm 2.5}$, collected on $7^{\rm th}$ July 2008. {\it Right panel: }Sampling of PM$_{\rm 2.5-10}$, collected on $27^{\rm th}$ October 2008.}
\label{fig:Aerosol_microscope}
\end{figure}

So far, all the techniques mentioned before provide information on aerosols as attenuators or scattering centres, but do not characterise the aerosols themselves. Such a characterization could help in the understanding of aerosol behaviour in the attenuation process. For instance, the atmospheric aerosols at the Pierre Auger Observatory are assumed to be desert-type particles~\cite{HAM_proceeding}. Also, in the air shower reconstruction done in the Auger Offline software, the aerosols are pure scatterers: absorption is not taken into account, since it is assumed to be very small. But this assumption depends on the chemical composition of the aerosols. The results from aerosol sampling add new information about the aerosols at the site and can be compared with measurements obtained by the instruments using optical techniques such as the CLF or the lidars. They can give a clearer idea about the origin of the aerosols present at the Auger Observatory, their sources and trajectories~(see Sect.~\ref{sec:hysplit}), and the connection between their composition and the meteorological variables in the region. An Andersen-Graseby 240 dichotomous sampler (see Fig.~\ref{fig:AndersenGraseby}~(left)) provided with polycarbonate membrane filters was used to separate fine size particles PM$_{\rm 2.5}$ ({\it i.e.}\ with an aerodynamic\footnote{{\it Aerodynamic diameter} is a physical property of a particle in a viscous fluid such as air. In general, particles have irregular shapes with actual geometric diameters that are difficult to measure. Aerodynamic diameter is an expression of a particle aerodynamic behaviour as if it were a perfect sphere having a density of 1~g/cm$^3$ and diameter equal to the aerodynamic diameter.} diameter $d\leq 2.5~\mu$m) and coarse particles PM$_{\rm 10}$ ($d\leq 10~\mu$m)~\cite{MariaI_LongPaper}. The sampling period was one day, beginning at 12:00~a.m. and ending at 12:00~a.m. of the following day. The samples were collected from June to November 2008, at the Coihueco location, on the roof of the FD building.

For gravimetric analysis and concentration measurements, a total of $34$~filters containing fine particles (PM$_{\rm 2.5}$) and $38$~filters containing coarse particles (PM$_{\rm 2.5-10}$) were considered. The concentrations, expressed as $\mu {\rm g/cm}^{3}$, were calculated as the ratio between the collected mass and the volume of air passed through the sampler during a full sampling period. Figure~\ref{fig:AndersenGraseby}~(right) gives the evolution in time of the measured aerosol mass concentrations. An increase of the aerosol concentration is observed from winter to spring, which could be related to decreasing snowfall and increasing temperature. Although snowfalls are rare during winter, the low temperatures keep the snow from melting for long periods, reducing atmospheric particulates near the ground.

\begin{table}[!t]
\centering
\begin{tabular}{c |  c c c c c c c c | c}
\hline\noalign{\smallskip}
\multicolumn{10}{c} {\bf Concentration fractions for each element} \\[0.5ex]
\noalign{\smallskip}\hline\noalign{\smallskip}
& Mg & Al & Si & P & S & K & Ca & Fe & Sum \\[0.5ex]

\hline
PM$_{\rm 2.5}~~~~$ [$\%$] & $3.5$ & $11.1$ & $60.0$ & & $1.2$ & & $6.3$ & & $82$\\
PM$_{\rm 2.5-10}$ [$\%$] & $4.4$ & $16.3$ & $69.4$ & $0.7$ & $0.3$ & $0.5$ & $4.6$ & $3.7$ & $100$\\ [1.0ex]
\noalign{\smallskip}\hline
\end{tabular}
~\\
\caption{\label {tab:aerosol_fractions}{\bf Element concentrations of PM$_{\rm \bf 2.5}$ and PM$_{\rm \bf 2.5-10}$ fractions, obtained by SEM/EDX (from~\cite{MariaI_LongPaper}).} Average particle concentrations, or atomic percentages, applied to two samples, one of each PM fraction.}
\end{table}	

Particle morphology and elemental composition were studied using a Scanning Electron Microscope (SEM) equipped with an Energy Dispersive X-ray system (EDX). Table~\ref{tab:aerosol_fractions} gives average compositions determined by SEM/EDX on at least $30$ individual particles. It shows that Si, Al, Ca, Mg, Fe and K, the typical mineral soil elements, are the major components. This indicates that aerosols are mainly dust suspended in the atmosphere. PM$_{\rm 2.5}$ fractions show more undetected particles due to the fact that beam focusing precision is insufficient to get a proper X-ray signal on particles smaller than one micron; thus many of them cannot be detected. Also, some particles in both fractions giving no detectable X-ray signal may be composed of light elements (Z $<$ 11), presumably organic matter. SEM micrographs of two representative PM$_{\rm 2.5}$ and PM$_{\rm 2.5-10}$ samples are shown in Fig.~\ref{fig:Aerosol_microscope}.

\section{Origin of atmospheric aerosols at Auger site using backward trajectory of air masses}
\label{sec:hysplit}
The Pierre Auger Collaboration has developed a large atmospheric monitoring program to have a better knowledge of aerosols present over its array. However, the aerosol population is highly mobile. Aerosol properties for a particular location are a function not only of local but also of emissions and meteorological conditions in the surrounding area. In our case, the Pierre Auger Observatory is affected by air masses potentially carrying aerosols from, for instance, Chile or the Pacific Ocean. Thus, any study having as a goal to explain aerosol properties over the Auger array needs to include work on the air mass origins. For a region outside of the Auger array to influence local measurements, the following three conditions must occur simultaneously: the region is a source of aerosols, the air mass coming from the region has to transport the aerosols in a sufficiently short time so that the aerosol content is still present on arrival, and the air mass arriving at the Auger array must be at an altitude that would affect the part of the atmosphere probed by the detector. Meteorological conditions also affect the aerosol transport to the observatory location.

\begin{figure}[!t]
\centering
\resizebox{1.0\textwidth}{!}{%
\includegraphics{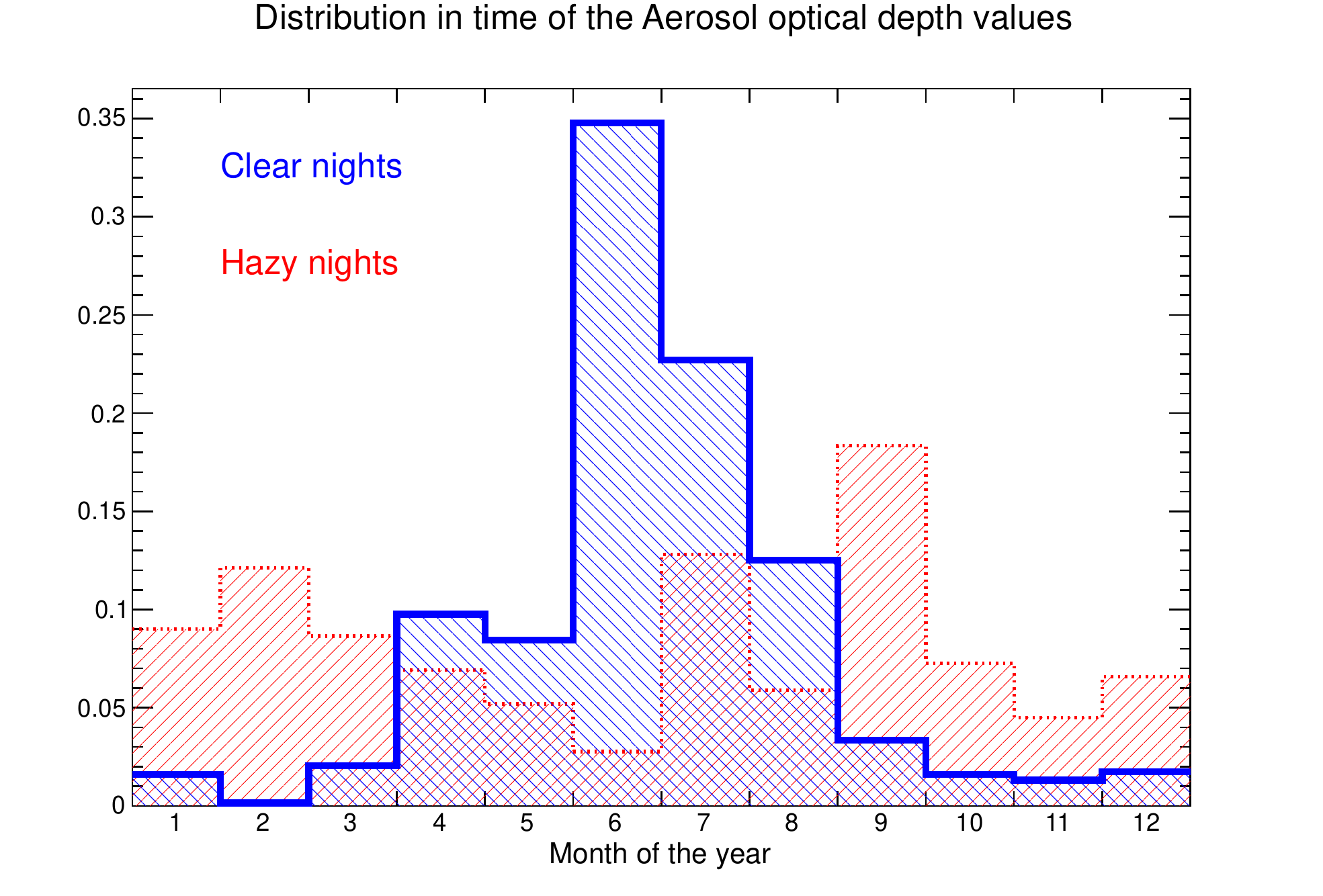}
}
\caption{{\bf Monthly frequency all along the year of clear and hazy nights.} Aerosol optical depths at 3.5~km above ground level measured between January 2004 and December 2010 at Los Morados. Clear nights are defined as $0.00\leq \tau_{\rm a}\leq 0.01$ {\it (continuous line)} and hazy nights as $\tau_{\rm a}\geq 0.10$ {\it (dotted line)}.}
\label{fig:VAOD_values_month}
\end{figure}

Different air models have been developed to study air mass relationships between two regions. Among them, the {\it HYbrid Single-Particle Lagrangian Integrated Trajectory} model, or HYSPLIT~\cite{HYSPLIT_1,HYSPLIT_2}, is a commonly used air-modelling program in atmospheric sciences that can calculate air mass displacements from one region to another. The HYSPLIT model developed by the Air Resources Laboratory, NOAA\footnote{NOAA, National Oceanic and Atmospheric Administration, U.S.A.}, is a complete system designed to support a wide range of simulations related to regional or long-range transport and dispersion of airborne particles. HYSPLIT computes simple trajectories to complex dispersion and deposition simulations using either puff or particle approaches with a Lagrangian framework. HYSPLIT can be used to get backward/forward trajectories: by moving backward/forward in time, the resulting backward/forward trajectory indicates air mass arriving at a receptor for a particular time, identifying the regions linked to it.

From values of aerosol optical depth given in Fig.~\ref{fig:CLF_VAOD_values} at Los Morados, the data sample is divided into three populations: the clear nights with the lowest aerosol concentrations ($\tau_{\rm a}\leq 0.01$), the hazy nights with the highest aerosol concentrations ($\tau_{\rm a}\geq 0.10$), the average nights being in the third category. Figure~\ref{fig:VAOD_values_month} shows the relative frequency month-by-month for clear nights and hazy nights. Clear nights are more common during the Austral winter than the rest of the year. Using the HYSPLIT tool, each clear or hazy night is assigned to a backward trajectory computed over 48 hours. Meteorological quantities used in calculations come from the GDAS model (Global Data Assimilation System), its validity having already been checked by the Pierre Auger Collaboration~\cite{GDASpaper}.

\begin{figure}[!t]
\centering
\resizebox{0.49\textwidth}{!}{%
\includegraphics{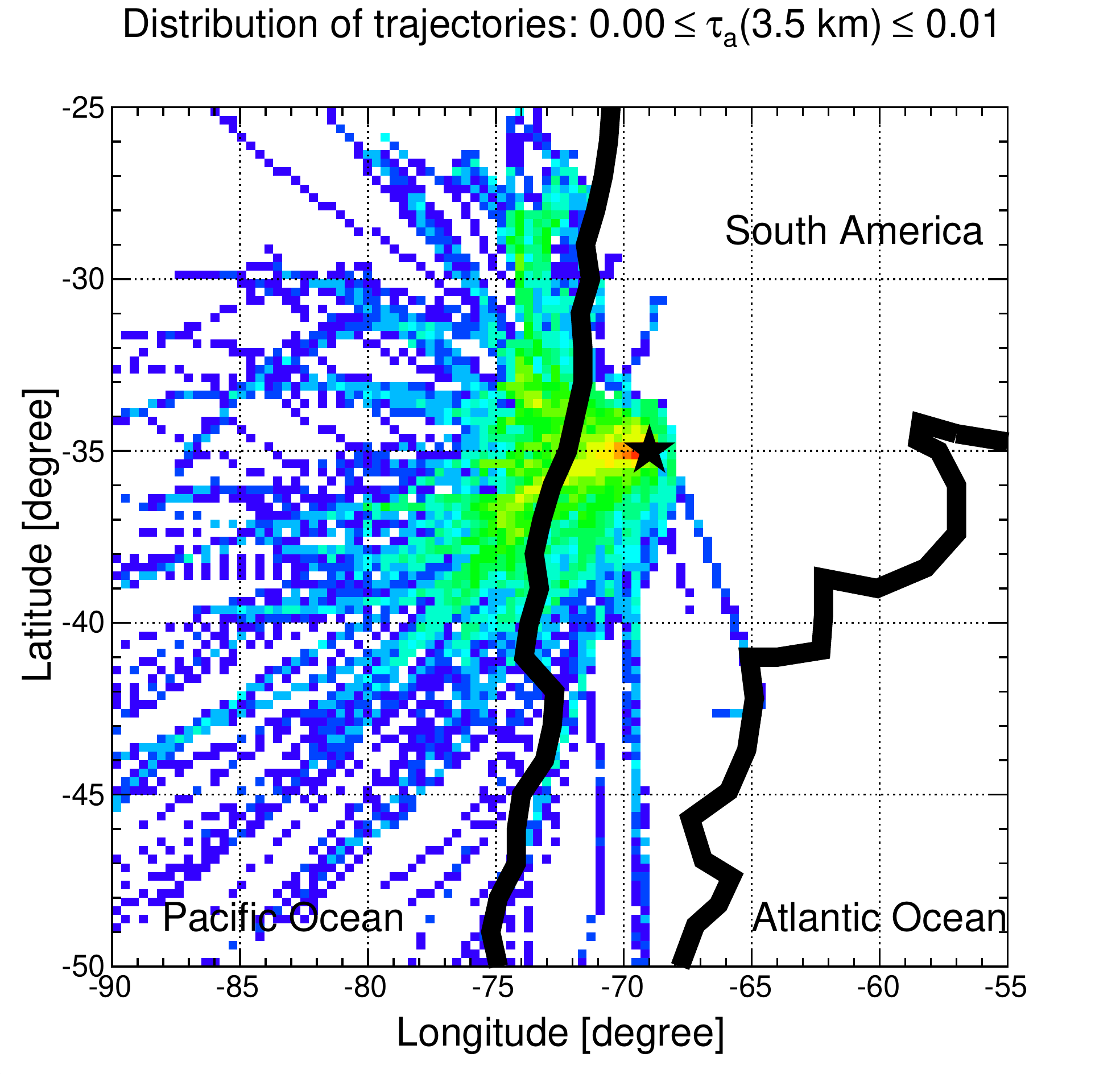}
}
\resizebox{0.49\textwidth}{!}{%
\includegraphics{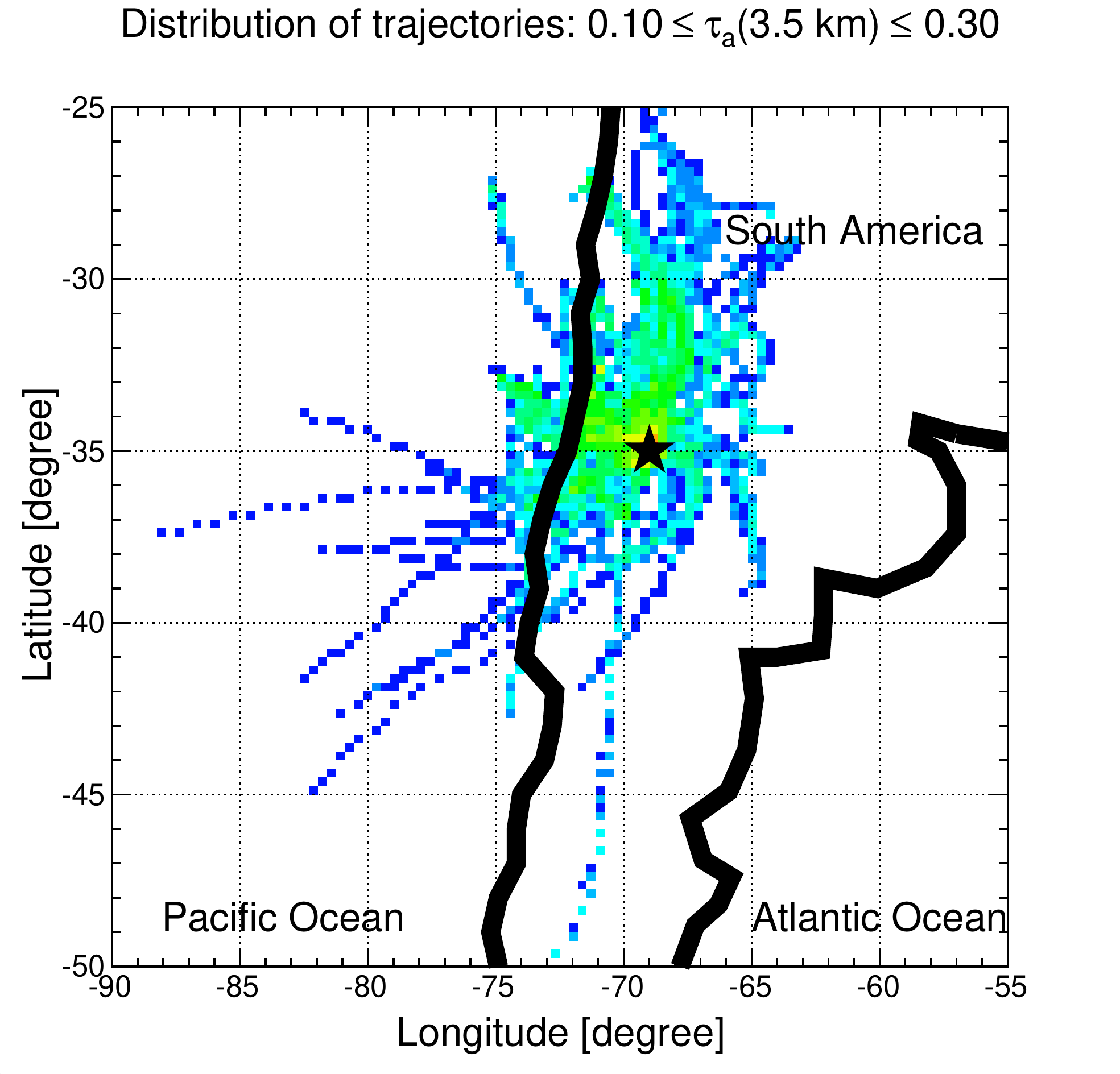}
}
\hspace{5cm}

\resizebox{0.49\textwidth}{!}{%
\includegraphics{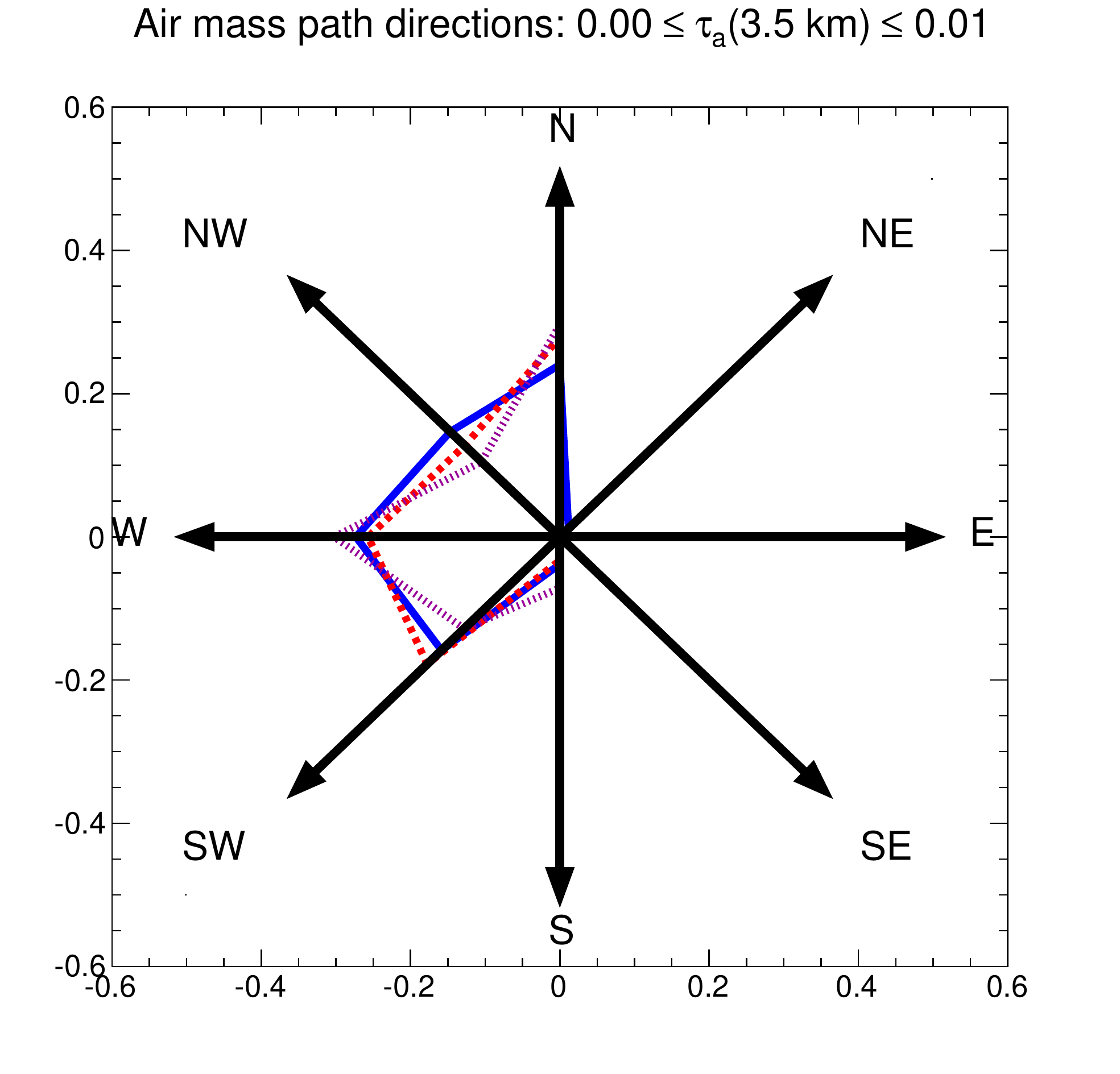}
}
\resizebox{0.49\textwidth}{!}{%
\includegraphics{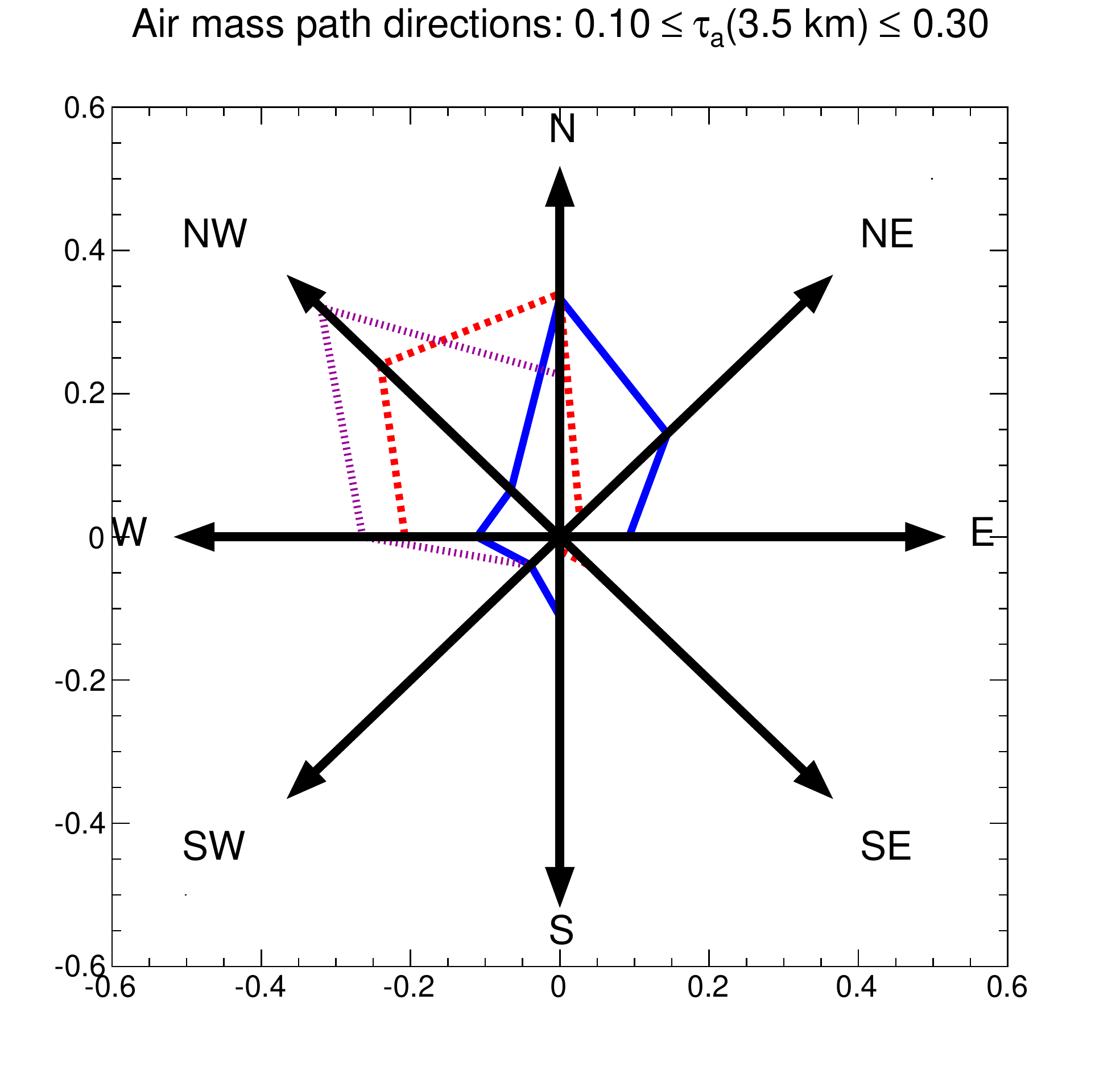}
}
\caption{{\bf Distribution of backward trajectories and direction of air masses for clear nights and hazy nights (from~\cite{KL_ECRS}).} Paths estimated with HYSPLIT for the years 2007/2008/2009 and aerosol optical depth data coming from the CLF measurements. 
{\it Top panel: }Distribution of 48-h backward trajectories from the Malarg\"ue location for a start altitude fixed at $500~$m AGL. 
{\it Bottom panel: }Direction of trajectories at three different start altitudes AGL: 500~m ({\it solid line in blue}), 1000~m ({\it dashed line in red}) and 3000~m ({\it dotted line in magenta}). Each distribution is normalised to one. Colours in online version. 
}
\label{fig:HYSPLIT_trajectories_VAOD}
\end{figure}

The distributions of the backward paths for clear nights and hazy nights are shown in Fig.~\ref{fig:HYSPLIT_trajectories_VAOD}~(top). The two plots are quite different: the air masses come mainly from the Pacific Ocean during the clear nights and travel principally through continental areas during the previous 48~hours for hazy nights. Following the conclusion of the chemical composition aerosol analysis presented in Sect.~\ref{sec:aerosol_sampling_measurements}, the aerosols originate mainly from the soil. Thus 48-h backward trajectories traveling mainly over the ocean can be characterised as air masses with a low concentration of soil aerosols. The aerosol contamination occurs only during the last hours for ``oceanic'' trajectories, whereas this is possible during the full period for the overland trajectories. It seems that backward trajectories are a key point, explaining the differences in aerosol concentration at the Pierre Auger Observatory. In Fig.~\ref{fig:HYSPLIT_trajectories_VAOD}~(bottom), directions of air masses are given for clear and hazy nights, for three different start altitudes at the location of the Pierre Auger Observatory: $500~$m AGL (blue diagram), $1000~$m AGL (red diagram) and $3000~$m AGL (magenta diagram). The same conclusions can be drawn for all three initial altitudes. Even if a slight shift to the West is seen for the highest altitude for hazy nights, it seems important to remind oneself that atmospheric aerosols are usually located in the low part of the atmosphere, typically in the first 2~km.

\section{The Pierre Auger Observatory as an aerosol observatory for Southern Hemisphere}
\label{sec:collab_inter}
We present in this section an interdisciplinary proposal for using the Pierre Auger Observatory as a scientific facility that will considerably enhance our knowledge of aerosol cycles over Southern Hemisphere. It has been shown in this text that the Pierre Auger Collaboration has developed a large atmospheric monitoring program with, currently, almost seven years of accumulated data~\cite{FP:Wiencke}. Beside the benefit for astroparticle measurements, a better knowledge of aerosol properties will enhance our capability of climate change forecasting because of the impact of dust deposition on oceanic biogeochemistry.  We would obtain a better accuracy on atmospheric fluorescence measurements plus valuable new data concerning continental dust emission, transport and deposition. The next sections explain the scientific issues of the proposed work.

\begin{figure}[!t]
\centering
\resizebox{1.\textwidth}{!}{%
\includegraphics{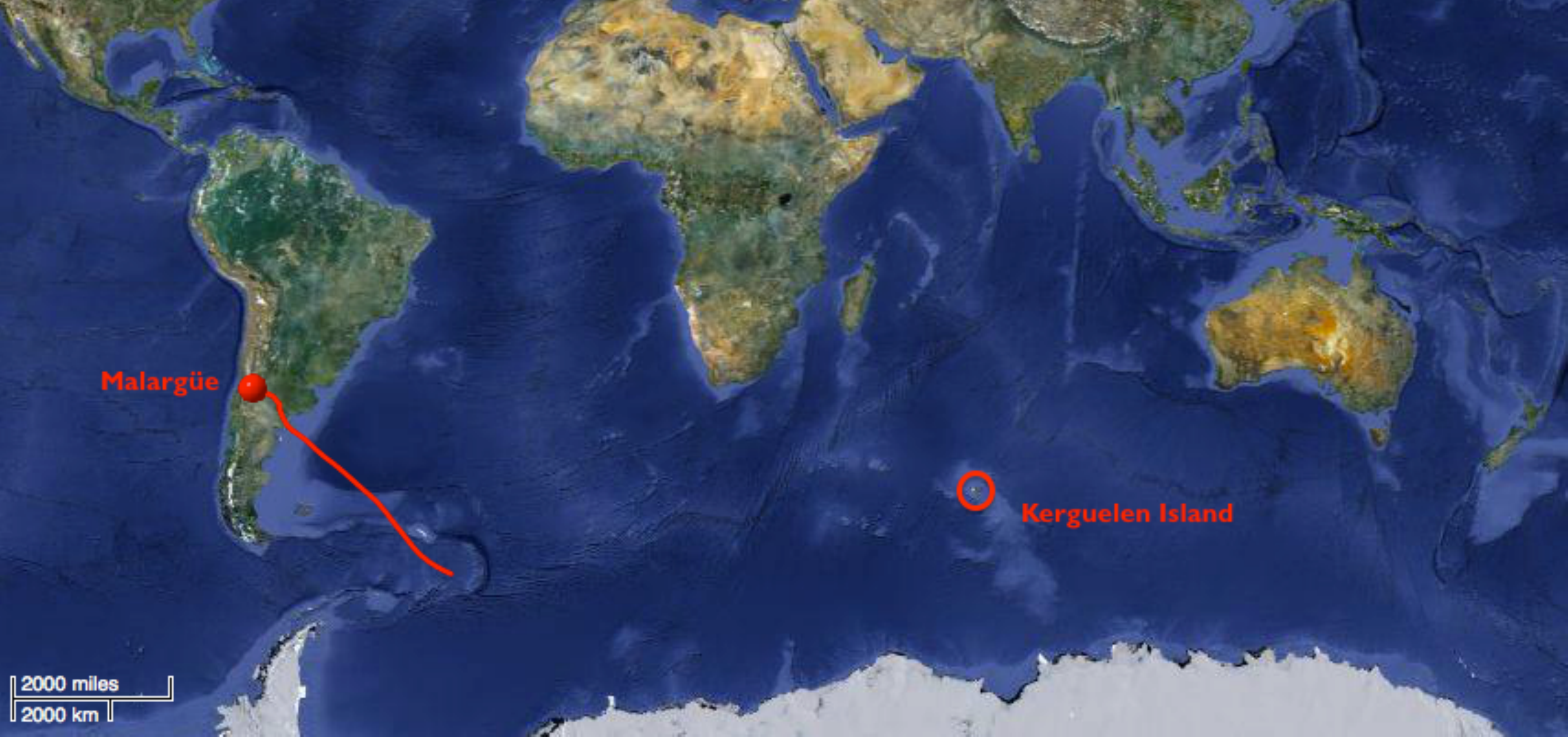}
}
\caption {{\bf Example of a forward trajectory from the Malarg\"ue location using HYSPLIT.} The initial height is fixed at $500~$m above ground level and the time scale is 100 hours (map taken from Google Earth). Kerguelen Island is represented with the circle.}
\label{fig:HYSPLIT_googlemap}
\end{figure}

\subsection{Impact of aerosol deposition on the oceans}
Atmospheric mineral aerosol, so-called ``dust'', is one of the major vectors feeding open ocean surface waters with trace metals~\cite{RL_2}. Even at extremely low concentrations, trace metals are micro-nutrients, necessary for the growth of phytoplankton~\cite{RL_2b}. In this way, trace metals are linked to climate change since they affect the capability of the marine biomass to trap CO$_2$ (see for example~\cite{RL_5}). The Austral region ranging from about 40$^\circ$ and 65$^\circ$ South is a major oceanic CO$_2$ sinks with exported carbon to deep sea water~\cite{RL_N_0}. This region is also very remote from continents and thus atmospheric dust exhibits very low concentrations~\cite{RL_8,RL_9,RL_12,RL_13}. This oceanic area is a HNLC region (High-Nutrients Low-Chlorophyll) and dust deposition could be a severe limiting factor for the primary production of nutrients~\cite{RL_1,RL_10}. Deposition measurements of continental dust have never been performed in this large area, except in the 1970's associated with radon~\cite{RL_7}. Some field measurements are available from point measurements at Crozet Island~\cite{RL_11}. Others use the deposition flux calculated from aerosol concentration measurements using rough assumptions based on {\it ad-hoc} deposition velocities or scavenging ratios~\cite{RL_13}. Recently, it has been shown by performing direct deposition measurements on Kerguelen Island that this last approach leads to atmospheric fluxes 50 to 100 times lower than those directly measured~\cite{RL_3}. There is an on-going debate on the importance of the atmospheric deposition flux over the Southern Ocean. Early modelling studies hypothesized a large flux, whereas {\it in-situ} measurements have led to a downward revision these early estimates. More recently, direct flux measurements again have pushed up the overall dust deposition flux to the Southern Ocean. Cassar {\it et al.}~\cite{RL_1} stated from modelling studies that the primary productivity is proportional to the aeolian dissolved iron deposition originated from dust. Iron bioavailability, and those of other nutrients, is conditioned by the ability of mineral dust to be dissolved in seawater. In the current estimation of atmospheric dissolved iron fluxes, the solubility of atmospheric iron in seawater remains a major source of uncertainty~\cite{RL_4}. Fluxes are generally calculated with an average solubility value obtained from available data in literature~\cite{RL_13,RL_1}. However the large majority of data are related only to African and Asian dust. Given the key role of the Austral Ocean on global climate (as a main oceanic CO$_2$ sinks), it is important to do studies that aimed to characterize iron and other trace metal solubility from Southern Hemisphere dust.

Since the Southern Hemisphere is mainly constituted of oceans, the only possible dust sources are South America including East Patagonia and East Andes Foothills, South Africa and Australia. South America is suspected to be the major dust source for the oceanic region ranging between 40$^\circ$~S and 60$^\circ$ S, with a major contribution of Patagonia but also with important sources located more North in the arid or semi-arid regions located East of the Andes mountains.

\subsection{Regional emission of mineral aerosol (dust)}
The Pierre Auger Observatory may be very useful for environmental science in general and particularly those concerning aerosols. The observatory is located in an area subjected to local emissions of soil dust but also to long-range transport of aerosols having marine, terrigeneous or anthropogenic origins. Such transported aerosols are often located at high altitudes and can be detected by laser beam scattering, in-situ aeroplane, kite or balloon sampling. They reach ground level during deposition processes. On the other hand, locally emitted aerosols are more present at ground level and can be well characterized by performing ground-based sampling on filters (see Sect.~\ref{sec:aerosol_sampling_measurements}). This should allow investigations of the chemical composition and particularly of the content in iron bearing particles that can absorb blue light~\cite{Lafon2006}. These aerosols can reach higher altitudes and be forward transported far away in the general sub-Antarctic atmospheric circulation (see Fig.~\ref{fig:HYSPLIT_googlemap}). Because of the sporadic character of aerosol emission pulses, continuous sampling should be carried out. Transportation models, including air mass trajectories, can provide a framework to understand the behaviour of the aerosol loading the atmosphere over the observatory, and assessing impacts on the Southern Ocean.

\subsection{Aerosol characterization and proposed action at the Pierre Auger Observatory}
In the previous sections, we have pointed out the influence of the aerosol and deposition chemistry on their potential impacts. Routine measurements are already being done using lidar systems, and particle number size distribution determined using optical particle counting. For mass balances, the particle mass size distribution will be deduced using model that makes assumptions about density and particle shape~\cite{RL_N_3}. Some preliminary sampling campaigns were done, using filtration or impaction. For filtration, air is pumped through a filter, often a membrane with a $0.2-0.4~\mu$m porosity. For impaction, a narrow air jet is directed against a plate, the overall geometry of the system will separate by impaction particles following their aerodynamical diameter. The larger particles impacting on the plate stay attached, whilst the smaller ones do not. If several of such impactor plates are cascaded, we obtain a so called cascade impactor which can deliver up to 13 size segregated levels of particles with a decreasing aerodynamical diameter. With a simple computation method, a continuous size distribution in mass can be rebuilt~\cite{RL_N_4}. Filters and plates must be brought to  laboratory where various chemical analyses are performed. Because of the high variability of atmospheric conditions and compositions, we propose to set up a continuous time series of aerosol sampling, allowing a quantitative annual budget. This continuous aerosol survey will run at Coihueco station, on a weekly basis. Various measurements will be done on filters: elemental analyses and more complex chemical analyses, including solubility. The only limitation is the preservation of the sample, which is not an issue except of contamination. {\it In-situ} experiments will use only small pumping systems and optical counters and do not need more than some hundreds of watts of electrical power. Most of time, work will be done at the laboratory.

For elemental analyses of collected samples, thin layer X-Ray fluorescence spectrometry is operated in laboratory without any transformation of the sample~\cite{RL_N_6,RL_N_7}, and at high rate and low cost. PIXE (Proton Induced X-Ray Emission) can also be used~\cite{RL_N_7,RL_N_8}. To access ultra-trace amounts and isotopic determinations, samples have to be digested in a mixture of various acids, to put all the sample in a soluble form. Elemental analyses are performed by Inductively Coupled Plasma Atomic Emission Spectrometry (ICP-AES)~\cite{RL_N_9} and Inductively Coupled Plasma Mass Spectrometry (ICP-MS)~\cite{RL_N_10}.

Elemental analyses will be performed on all the collected samples for major and trace elements using XRF. These first analyses will be used to select interesting samples. On these samples, ultra-trace and solubility measurements, which require more work, will be done using ICP-AES and ICP-MS. For solubility, water at various pH and various composition, including sea water, is passed through the sample and dissolves parts of the aerosols in a kinetic experiment~\cite{RL_N_11,RL_N_12}. Variations of dissolution rate along time but also acidity of the water give valuable information on the bioavailability of the collected aerosols but also chemical bonding in the solid aerosol. To enhance the interpretation of such chemical data, particles will be directly observed by Transmission Electronic Microscopy (TEM) and Scattered Electronic Microscopy (SEM). Microscopy is devoted to single particle analyses, and observation of numerous particles can give very detailed information on size and shape distribution, chemical composition and mineralogy of the collected aerosols~\cite{RL_N_13}. Such information, together with bulk elemental analyses and solubility measurements are valuable for inclusion in a global model of dust solubility and dust impact on biogeochemical cycles~\cite{RL_N_14}. Such knowledge will be finally used to enhance lidar and CLF measurements interpretations and to improve our knowledge of the influence of aerosol chemical composition on the astroparticles atmospheric signals. Indeed, for instance, it would be interesting to check if assumptions on aerosol absorption done up to now by the Pierre Auger Collaboration are validated.

\section*{Conclusion}
The Pierre Auger Collaboration has accumulated a large database of aerosol measurements. This effort has significantly reduced systematic uncertainties in the cosmic ray air shower reconstruction. Cross-checks between optical measurements and sampling, including their size distribution or their chemical composition, are now being undertaken at the observatory to improve our knowledge on aerosols.

Collaborative works with scientists from the environmental sciences are currently discussed in the Pierre Auger Collaboration. Indeed, the observatory is located in a region where only few atmospheric survey stations are installed. With additional facilities as enumerated in  Sect.~\ref{sec:collab_inter}, the Pierre Auger Observatory could contribute to a better knowledge of the Southern Hemisphere atmosphere and its chemical composition. The observatory could be candidate to become member of international networks as the Global Atmosphere Watch (GAW) program~\cite{GAW} or the Aerosol Robotic Network (AERONET)~\cite{AERONET} and more generally to be a reference station.

The proposed collaborative project would considerably enhance knowledge of the optical and chemical properties of aerosols over the Pierre Auger Observatory but also the global impact of airborne particles over Southern Hemisphere.

\section*{Acknowledgements}
The authors would like to thank especially their collaborators Roger Clay, Vincenzo Rizi, Darko Veberi\v{c}, Lawrence Wiencke and Martin Will for their comments and improvements made as a result. One of the authors, Karim Louedec, thanks Dr Marcel Urban, who was in France at the forefront of promoting interdisciplinary science for the Pierre Auger Observatory and one of its greatest proponents.

\end{document}